# Thermodynamics of amide + ketone mixtures. 2. Volumetric, speed of sound and refractive index data for *N,N*-dimethylacetamide + 2-alkanone systems at several temperatures. Application of Flory's model to tertiary amide + *n*-alkanone systems

ANA COBOS[(1)], JUAN ANTONIO GONZÁLEZ*[(1)], FERNANDO HEVIA,[(1)] ISAÍAS GARCÍA DE LA FUENTE[(1)] AND CRISTINA ALONSO TRISTÁN[(2)]

[(1)] Dpto. Ingeniería Electromecánica. Escuela Politécnica Superior. Avda. Cantabria s/n. 09006 Burgos, (Spain)

[(2)] G.E.T.E.F., Departamento de Física Aplicada, Facultad de Ciencias, Universidad de Valladolid, Paseo de Belén, 7, 47011 Valladolid, Spain,

*e-mail: jagl@termo.uva.es; Fax: +34-983-423136; Tel: +34-983-423757

**Abstract**

Data on density, $\rho$, speed of sound, *c*, and refractive index, $n_D$, have been reported at (293-303.15) K for the *N,N*-dimethylacetamide (DMA)+ $CH_3CO(CH_2)_{u-1}CH_3$ (*u* = 1,2,3) systems, and at 298.15 K for the mixture with *u* = 5. These data have been used to compute excess molar volumes, $V_m^E$, excess adiabatic compressibilities, $\kappa_S^E$, and excess speeds of sound $c^E$. Negative $V_m^E$ values indicate the existence of structural effects and interactions between unlike molecules. From molar excess enthalpies, $H_m^E$, available in the literature for *N,N*-dimethylformamide (DMF), or *N*-methylpyrrolidone (NMP) + *n*-alkanone systems, it is shown: (i) amide-ketone interactions are stronger in DMF systems than in those with NMP; (ii) they become weaker when *u* increases in mixtures with a given amide. Structural effects largely contribute to $H_m^E$ and are more relevant in mixtures containing NMP. The application of the Flory's model reveals that the random mixing hypothesis is valid in large extent for DMF solutions, while NMP systems are characterized by rather strong orientational effects. From values of molar refraction and of the product $P_{int} V_m$ (where $P_{int}$ is the internal pressure and $V_m$ the molar volume), it is concluded that dispersive interactions increase with *u*, or when DMF is replaced by DMA in mixtures with a fixed ketone.

**KEYWORDS: amides; *n*-alkanones; thermophysical properties; Flory; random mixing**

## 1. Introduction

Tertiary amides are aprotic solvents of high polarity, with very good donor-acceptor properties, capable to solve many organic substances. They have many applications as selective extractants of aromatic and saturated hydrocarbons and of nitrogen compounds in the fuel industry [1, 2]. These amides are also useful for the production of acrylic or fibers, plastics, pesticides or surface coatings, in nanotechnology [3-5], or in pharmaceutical industry [6, 7]. In addition, the research of liquid mixtures containing the amide group is suitable for a deeper understanding of complex molecules of biological interest [8]. For example, the water + *N,N*-dimethylformamide (DMF) system may be considered as a simple biochemical model of biological aqueous solutions [9,10]. On the other hand, the significant local order shown by pure amides makes also very interesting their theoretical study [11]. In the case of *N,N*-dialkylamides, due to the lack of hydrogen-bonds, this has been attributed to the existence of strong dipolar interactions [10,12].

2-Alkanones, ($CH_3CO(CH_2)_{u-1}CH_3$) are polar, aprotic compounds and hydrogen bonds acceptors, used as solvents for plastics and some synthetic fibers or as intermediates in the obtention of very important compounds (e.g., methyl methacrylate). They have also an essential role in biochemistry, as many sugars are ketones, and fatty acid synthesis proceeds via these compounds. These features make suitable the study of amide + alkanone mixtures, which must be taken into account within a general experimental and theoretical research of liquid mixtures containing the functional groups carbonyl, amine or amide. In a previous work, we have provided data on density, $\rho$, speeds of sound, $c$, and refractive indices, $n_D$, at (293.15-303.15) K for DMF systems with $u$ = 1,2,3 and at 298.15 K for the solution with $u$ = 5 [13]. As a continuation of this research, we report now similar data for binary mixtures containing *N,N*-dimethylacetamide (DMA) and the same 2-alkanones over the same temperature range. In order to complete our study, data available in the literature on *N*-methylpyrrolidone (NMP) + *n*-alkanone systems are also considered [14-17]. This allows examine the dependence of thermophysical properties on the size and shape of the amide in systems with a given *n*-alkanone, or on the chain length of the *n*-alkanone in mixtures with a given amide. Finally, amide + *n*-alkanone mixtures are treated in terms of the Flory model [18-20], in order to explore the validity of the random mixing hypothesis for such solutions. We have widely used this approach to investigate orientational effects in liquid mixtures, such as 1-alkanol + linear or cyclic monoether [20], or + polyether [21], or ether + benzene, or + toluene [22].

## 2. Experimental

### *2.1 Materials*

Information regarding the source and the purity of the compounds considered is shown in Table 1. All the chemicals have been used without further purification. Table 2 lists experimental values of $\rho$, $c$, $n_{\mathrm{D}}$, thermal expansion coefficient, $\alpha_p$, isentropic compressibility, $\kappa_S$, and isothermal compressibility, $\kappa_T$, for the pure compounds. Our values are in good agreement with data from the literature.

*2.2 Apparatus and procedure*

Binary mixtures have been prepared by mass in small vessels ($\approx 10$ $cm^3$) using an analytical balance Sartorius MSU125p (weighing accuracy 0.01 mg), with all weighings corrected for buoyancy effects. The standard uncertainty in the final mole fraction is estimated to be 0.0001. Molar quantities were determined on the basis of the relative atomic mass Table of 2015 issued by the CIAAW (IUPAC) [23]. Temperatures were measured by means of Pt-100 resistances. Their calibration was conducted according to the ITS-90 scale of temperature, against the triple point of the water and the fusion point of Ga. The standard uncertainty of the equilibrium temperature measurements is 0.01 K and 0.02 K. for $\rho$ and $n_{\mathrm{D}}$ measurements, respectively. A vibrating-tube densimeter and sound analyser DSA5000 from Anton Paar, automatically thermostated within $0.01\,\mathrm{K}$, has been used for the measurements of $\rho$ and $c$ values. Details about the device calibration can be found elsewhere [24]. The repeatability of the $\rho$ measurements is $0.005\,\mathrm{kg{\cdot}m^{-3}}$, whereas their overall uncertainty is $1 \cdot 10^{-2}$ $\mathrm{kg{\cdot}m^{-3}}$. Speed of sound is determined measuring the time of propagation of short acoustic pulses (central frequency, 3 MHz) [25], transmitted repeatedly through the sample. The repeatability of the $c$ measurements is $0.1\,\mathrm{m{\cdot}s^{-1}}$ and their standard uncertainty is $0.2\,\mathrm{m{\cdot}s^{-1}}$.

The molar excess volume, $V_{\mathrm{m}}^{\mathrm{E}}$, and the excess speed of sound, $c^{\mathrm{E}}$, of the system cyclohexane + benzene have been measured at (293.15-303.15) K to check the experimental technique. Our experimental results are in good agreement with published values [26-28]. The standard uncertainty of $V_{\mathrm{m}}^{\mathrm{E}}$ is $(0.010\left|V_{\mathrm{m,max}}^{\mathrm{E}}\right| + 0.005)$ $\mathrm{cm^3{\cdot}mol^{-1}}$, where $\left|V_{\mathrm{m,max}}^{\mathrm{E}}\right|$ stands for the maximum absolute experimental value of $V_{\mathrm{m}}^{\mathrm{E}}$ respect to the composition. The standard uncertainty of $c^{\mathrm{E}}$ is estimated to be $0.4\,\mathrm{m{\cdot}s^{-1}}$. A refractometer RFM970 from Bellingham+Stanley has been used for the $n_{\mathrm{D}}$ measurements. The technique is based on the optical detection of the critical angle at the wavelength of the sodium D line (589.3 nm). The temperature is controlled by means of Peltier modules and its stability is $0.02\,\mathrm{K}$. The refractometer was calibrated using 2,2,4-trimethylpentane and toluene at (293.15-303.15) K, as recommended by Marsh [29]. The repeatability of the measurements is $0.00004$, and the standard uncertainty is $0.00008$.

## 3. Equations

The Anton Paar DSA5000 apparatus allows to obtain rather directly experimental values of $\rho$, molar volume, $V_{\mathrm{m}}$, $\alpha_p$, and $\kappa_S$. The values of $\alpha_p = -\frac{1}{\rho}\left(\frac{\partial \rho}{\partial T}\right)_p$ have been determined assuming that $\rho$ depends linearly on $T$ in the range of temperatures considered. In addition, as long as it is possible to neglect the dispersion and absorption of the acoustic wave, $\kappa_S$ can be obtained using the Newton-Laplace equation:

$$\kappa_S = \frac{1}{\rho c^2} \tag{1}$$

The values $F^{\mathrm{id}}$ of a magnitude, $F$, for an ideal mixture at the same temperature and pressure as the investigated solution are calculated from the relations [30-32]:

$$F^{\mathrm{id}} = x_1 F_1 + x_2 F_2 \qquad \left(F = V_{\mathrm{m}}, C_{p\mathrm{m}}\right) \tag{2}$$

$$F^{\mathrm{id}} = \phi_1 F_1 + \phi_2 F_2 \qquad \left(F = \alpha_p, \kappa_T\right) \tag{3}$$

where $F_i^*$ denotes the property for the pure component $i$, $C_{p\mathrm{m}}$ is the molar heat capacity at constant pressure, and $\phi_i = x_1 V_{\mathrm{m}i} / V_m^{\mathrm{id}}$ represents the ideal volume fraction. For $\kappa_S$ and $c$, the following expressions are used [30]:

$$\kappa_S^{\mathrm{id}} = \kappa_T^{\mathrm{id}} - \frac{T V_{\mathrm{m}}^{\mathrm{id}} \left(\alpha_p^{\mathrm{id}}\right)^2}{C_{p\mathrm{m}}^{\mathrm{id}}} \tag{4}$$

$$c^{\mathrm{id}} = \left(\frac{1}{\rho^{\mathrm{id}} \kappa_S^{\mathrm{id}}}\right)^{1/2} \tag{5}$$

being $\rho^{\mathrm{id}} = \left(x_1 M_1 + x_2 M_2\right) / V_{\mathrm{m}}^{\mathrm{id}}$ the ideal density, and $M_i$ the molar mass of the pure component $i$. The ideal values of $n_{\mathrm{D}}$ are obtained from [33]:

$$n_{\mathrm{D}}^{\mathrm{id}} = \left[\phi_1 \left(n_{\mathrm{D}1}\right)^2 + \phi_2 \left(n_{\mathrm{D}2}\right)^2\right]^{1/2} \tag{6}$$

The excess properties, $F^{\mathrm{E}}$, are then obtained using the relation:

$$F^{\mathrm{E}} = F - F^{\mathrm{id}} \qquad \left(F = V_{\mathrm{m}}, \kappa_S, c, \alpha_p, n_{\mathrm{D}}\right) \tag{7}$$

## 4. Experimental results

Values of $\rho$, $c$, and $V_{\mathrm{m}}^{\mathrm{E}}$, at the considered temperatures, as functions of $x_1$, the mole fraction of DMA, are included in Table 3 (Figure 1). For the 2-heptanone mixture, the measurements were made at 298.15 K only due to: (i) the very low $|V_{\mathrm{m}}^{\mathrm{E}}|$ values encountered for this system; (ii) the weak temperature dependence of $V_{\mathrm{m}}^{\mathrm{E}}$ encountered for the remaining mixtures. The corresponding results of $\kappa_S^{\mathrm{E}}$, $c^{\mathrm{E}}$, and $\alpha_p^{\mathrm{E}}$ at 298.15 K are given in Table 4 (Figures 2, 3). The $n_{\mathrm{D}}$ values and their corresponding excess functions, $n_{\mathrm{D}}^{\mathrm{E}}$, are collected in Table 5 (Figure 4).

The data have been fitted by an unweighted linear least-squares regression to a Redlich-Kister equation:

$$F^{\mathrm{E}} = x_1\left(1 - x_1\right)\sum_{i=0}^{k-1} A_i\left(2x_1 - 1\right)^i \qquad \left(F = V_{\mathrm{m}}, \kappa_S, c, \alpha_p, n_{\mathrm{D}}\right) \tag{8}$$

For each system and property, the number, $k$, of necessary coefficients for the regression has been determined by applying an F-test of additional term [34] at 99.5% confidence level. Table 6 includes the parameters $A_i$ obtained, and the standard deviations $\sigma\left(F^{\mathrm{E}}\right)$, defined by:

$$\sigma\left(F^{\mathrm{E}}\right) = \left[\frac{1}{N-k}\sum_{j=1}^{N}\left(F_{\mathrm{cal},j}^{\mathrm{E}} - F_{\mathrm{exp},j}^{\mathrm{E}}\right)^2\right]^{1/2} \tag{9}$$

where the index $j$ = 1, *N* (number of experimental data $F_{\mathrm{exp},j}^{\mathrm{E}}$), and $F_{\mathrm{cal},j}^{\mathrm{E}}$ is the corresponding value of the excess property $F^{\mathrm{E}}$ calculated from equation (8).

## 5. Flory model

### *5.1. Hypotheses and equations*

A short summary of the main hypotheses of the theory foloows [18,19,35-37]. (i) Molecules are formed by segments, which are arbitrarily chosen isomeric portions of a molecule. (ii) The mean intermolecular energy per contact is assumed to be proportional to $-\eta / v_s$ (where $\eta$ (>0) is a constant characterizing the energy of interaction for a pair of neighbouring sites and $v_s$ is the segment volume). (iii) Restrictions on the precise location of a given segment by its neighbours in the same chain are taken into account, when the configurational partition function is determined, assuming that the number of external degrees of freedom of the segments is lower than 3. (iv) Random mixing is assumed. The probability of having species of kind *i* neighbours to any given site is equal to the site fraction ($\theta_i$). If the total number of contact sites is very large, the probability of formation of an interaction between contacts sites belonging to different liquids is $\theta_1\theta_2$. Under these hypotheses, the Flory equation of state is:

$$\frac{\bar{P}\bar{V}}{\bar{T}} = \frac{\bar{V}^{1/3}}{\bar{V}^{1/3}-1} - \frac{1}{\bar{V}\bar{T}} \tag{10}$$

where $\bar{V} = V/V^*$; $\bar{P} = P/P^*$ and $\bar{T} = T/T^*$ are the reduced volume, pressure and temperature, respectively. Equation (10) is valid for pure liquids and liquid mixtures. For pure liquids, the reduction parameters, $V_i^*$, $P_i^*$ and $T_i^*$ are obtained from data on $\rho_\mathrm{i}$, $\alpha_{P\mathrm{i}}$, $\kappa_{T\mathrm{i}}$. The corresponding expressions for reduction parameters for mixtures are given elsewhere [20]. $H_\mathrm{m}^\mathrm{E}$ is determined from,

$$H_\mathrm{m}^\mathrm{E} = \frac{x_1 V_1^* \theta_2 X_{12}}{\bar{V}} + x_1 V_1^* P_1^* \left(\frac{1}{\bar{V}_1} - \frac{1}{\bar{V}}\right) + x_2 V_2^* P_2^* \left(\frac{1}{\bar{V}_2} - \frac{1}{\bar{V}}\right) \tag{11}$$

All the symbols have their usual meaning [22]. In equation (11), the term which depends directly on $X_{12}$ is the interaction contribution to $H_\mathrm{m}^\mathrm{E}$. The remaining terms are the so-called equation of state contribution to $H_\mathrm{m}^\mathrm{E}$. The reduced volume of the mixture, $\bar{V}$, in equation (11) is obtained from the equation of state. Therefore, the molar excess volume can be also calculated:

$$V_\mathrm{m}^\mathrm{E} = (x_1 V_1^* + x_2 V_2^*)(\bar{V} - \varphi_1 \bar{V}_1 - \varphi_2 \bar{V}_2) \tag{12}$$

### 5.2 *Estimation of the Flory interaction parameter*

$X_{12}$ is determined from a $H_\mathrm{m}^\mathrm{E}$ measurement at given composition from [20-22]:

$$X_{12} = \frac{x_1 P_1^* V_1^* \left(1 - \frac{\bar{T}_1}{\bar{T}}\right) + x_2 P_2^* V_2^* \left(1 - \frac{\bar{T}_2}{\bar{T}}\right)}{x_1 V_1^* \theta_2} \tag{13}$$

For the application of this expression, we note that $\overline{VT}$ is a function of $H_{\mathrm{m}}^{\mathrm{E}}$:

$$H_{\mathrm{m}}^{\mathrm{E}} = \frac{x_1 P_1^* V_1^*}{\overline{V}_1} + \frac{x_2 P_2^* V_2^*}{\overline{V}_2} + \frac{1}{\overline{VT}}(x_1 P_1^* V_1^* \overline{T}_1 + x_2 P_2^* V_2^* \overline{T}_2) \tag{14}$$

and that from the equation of state, $\overline{V} = \overline{V}(\overline{T})$. More details have been given elsewhere [20-22]. Equation (13) generalizes of that previously given to calculate $X_{12}$ from $H_{\mathrm{m}}^{\mathrm{E}}$ at $x_1$ = 0.5 [38] and allows to investigate the concentration dependence of $X_{12}$. Systems characterized by a behaviour close to that of random mixing show a weak concentration dependence of $X_{12}$ [21,22]. Properties of tertiary amides and *n*-alkanones molar volumes, $V_{\mathrm{i}}$, $\alpha_{Pi}, \kappa_{Ti}$, and the corresponding reduction parameters, $P_i^*$ and $V_i^*$, needed for calculations are collected in Table S1 (supplementary material). Table 7 lists $X_{12}$ values determined from experimental $H_{\mathrm{m}}^{\mathrm{E}}$ data at $x_1$ = 0.5.

*5.3 Results*

Results on $H_{\mathrm{m}}^{\mathrm{E}}$ obtained from the Flory model are collected in Table 7 (see Figures 5-6). This Table includes the interactional contribution to $H_{\mathrm{m}}^{\mathrm{E}}$, $H_{\mathrm{m,int}}^{\mathrm{E}}$, and the relative standard deviations for $H_{\mathrm{m}}^{\mathrm{E}}$ defined as:

$$\sigma_r(H_{\mathrm{m}}^{\mathrm{E}}) = \left[\frac{1}{N}\sum\left(\frac{H_{\mathrm{m,exp}}^{\mathrm{E}} - H_{\mathrm{m,calc}}^{\mathrm{E}}}{H_{\mathrm{m,exp}}^{\mathrm{E}}}\right)^2\right]^{1/2} \tag{15}$$

where *N* (=19) is the number of data points, and $H_{\mathrm{m,exp}}^{\mathrm{E}}$ stands for the smoothed $H_{\mathrm{m}}^{\mathrm{E}}$ values calculated at $\Delta x_1$ = 0.05 in the composition range [0.05, 0.95] from polynomial expansions, previously checked, given in the original works. Figure S1 (supplementary material) shows the concentration dependence of $X_{12}$ for some systems.

## 6. Discussion

Hereafter, we are referring to values of the excess functions and of thermophysical properties at $T = 298.15\mathrm{K}$ and equimolar composition.

The amides and ketones considered along the work are very polar compounds. Thus, the dipole moment, $\mu$, of the amides are (in D): 3.7 (DMF); 3.81 (DMA) [39]; 4.09 (NMP) [40]. For 2-alkanones, $\mu$/D = 2.69 (acetone); 2.76 (2-butanone); 2.70 (2-pentanone); 2.59 (2-heptanone) [40]. However, the impact of polarity on bulk properties is better evaluated by means of the effective dipole moment, $\overline{\mu}$ [41-43]:

$$\bar{\mu} = \left[ \frac{\mu^2 N_A}{4\pi\varepsilon_0 V_m \kappa_B T} \right]^{1/2} \tag{16}$$

where $N_A$, $\varepsilon_0$, $\kappa_B$ stand for the Avogadro's number, the permittivity of the vacuum, and the Boltzmann's constant, respectively. The $\bar{\mu}$ values are: 1.60 (DMF); 1.51 (DMA); 1.59 (NMP); 1.20 (2-propanone); 1.11 (2-butanone); 1.00 (2-pentanone); 0.83 (2-heptanone). It is to be noted that, for a given homologous series (2-alkanones), $\mu$ varies only slightly with the chain length of the compound, while the $\bar{\mu}$ variation is much greater.

Mixtures of tertiary amides with *n*-alkanes show miscibility gaps at 298.15 K. In fact, the upper critical solution temperature, UCST/K, of heptane systems changes in the order: 342.55 (DMF) [44] > 326.05 (NMP) [45] > 309.8 (DMA) [46]. This reveals that dipolar interactions between amide molecules become weaker in the same sequence. Results for excess molar enthalpies at infinite dilution of the amide, $H_{\mathrm{m,1}}^{\mathrm{E},\infty}$ in systems with heptane or cyclohexane at 298.15 K also support this conclusion as it is indicated by the following values (in kJ·mol$^{-1}$): 17.1 (DMF + heptane) [47]; 12.0 (NMP + heptane); 11.7 (NMP + cyclohexane) [48]. If one takes into account the similar $\bar{\mu}$ values for DMF ($V_{\mathrm{m}}$=77.42 cm$^3$·mol$^{-1}$ [13]) and NMP ($V_{\mathrm{m}}$= 96.63 cm$^3$·mol$^{-1}$ [40]), this means that dipolar interactions are not only determined by $\bar{\mu}$ values. Effects related to the size and shape of the molecules must also be considered. In fact, the rather large difference between UCST values of mixtures with NMP or DMA ($V_{\mathrm{m}}$= 93.04 cm$^3$·mol$^{-1}$) should be ascribed essentially to the different shape of these amides.

2-Alkanone + *n*-alkane systems are characterized by large and positive $H_{\mathrm{m}}^{\mathrm{E}}$ values, accordingly with their rather large $\bar{\mu}$ values. For example, $H_{\mathrm{m}}^{\mathrm{E}}$(heptane)/J·mol$^{-1}$ = 1704 (2-propanone) [49]; 1339 (2-butanone) [50]; 886 (2-heptanone) [51]. We note that $H_{\mathrm{m}}^{\mathrm{E}}$ decreases with the decreasing of $\bar{\mu}$, indicating that the observed $H_{\mathrm{m}}^{\mathrm{E}}$ variation is related to a weakening of the dipolar interactions between ketone molecules [52]. This also leads to a lower contribution to $V_{\mathrm{m}}^{\mathrm{E}}$ from the disruption of ketone-ketone interactions and to decreased $V_{\mathrm{m}}^{\mathrm{E}}$ (heptane)/cm$^3$·mol$^{-1}$ values in the sequence: 1.129 (acetone) [53] > 0.280 (2-octanone) [54].

### *6.1 Calorimetric data*

Tertiary amide + *n*-alkanone mixtures are characterized by low and even negative $H_{\mathrm{m}}^{\mathrm{E}}$ values. Thus, $H_{\mathrm{m}}^{\mathrm{E}}$(DMF)/J·mol$^{-1}$ = 34 (2-propanone) [16]; 142 (2-butanone), 206 (2-pentanone) [17], and for NMP mixtures: – 165 (2-butanone) [14]; 226 (4-heptanone); 362 (5-nonanone) [15]. On the basis of the features previously discussed for tertiary amide or *n*-alkanone + alkane mixtures, these low $H_{\mathrm{m}}^{\mathrm{E}}$ values underline the existence of strong amide-alkanone interactions in the systems under investigation. In a previous work [13], we evaluated the enthalpy of the mentioned interactions, $\Delta H_{\text{NCO-CO}}$, for DMF solutions. The procedure applied is now briefly explained. It is well known that if structural effects are neglected [43,55], $H_{\mathrm{m}}^{\mathrm{E}}$ is the result of three contributions: two of them are positive, $\Delta H_{\text{NCO-NCO}}, \Delta H_{\text{CO-CO}}$, and are related to the breaking of amide-amide and ketone-ketone interactions upon mixing, respectively; $\Delta H_{\text{NCO-CO}}$ is a negative contribution due to the new NCO---CO interactions created along the mixing process. That is [56-58]:

$$H_{\mathrm{m}}^{\mathrm{E}} = \Delta H_{\text{NCO-NCO}} + \Delta H_{\text{CO-CO}} + \Delta H_{\text{NCO-CO}} \qquad (17)$$

Equation (17) may be extended to $x_1 \to 0$ [56,58] to evaluate $\Delta H_{\text{NCO-CO}}$. In such a case, $\Delta H_{\text{NCO-NCO}}$ and $\Delta H_{\text{CO-CO}}$ can be replaced by $H_{\mathrm{m1}}^{\mathrm{E},\infty}$ (partial excess molar enthalpy at infinite dilution of the first component) of tertiary amide or *n*-alkanone + heptane systems. Thus,

$$\Delta H_{\text{NCO-CO}} = H_{\mathrm{m1}}^{\mathrm{E},\infty}(\text{amide } + n\text{-alkanone})$$

$$-H_{\mathrm{m1}}^{\mathrm{E},\infty}(\text{amide } + \text{heptane}) - H_{\mathrm{m1}}^{\mathrm{E},\infty}(n\text{-alkanone} + \text{heptane}) \qquad (18)$$

The values (in kJ·mol$^{-1}$) determined for DMF systems were [13]: $-26.0$ (2-propanone); $-24.0$ (2-butanone); $-22.6$ (2-pentanone). These results show: (i) DMF-2-alkanone interactions are rather strong. Note that the value typically used in the ERAS model for the enthalpy of the self-association of 1-alkanols is $-25.1$ kJ·mol$^{-1}$ [41, 59-61]; (ii) the enthalpy of the mentioned interactions become weaker when *u* is increased, probably due to the carbonyl group becomes then more sterically hindered. For NMP systems, we have obtained (Table S2, supplementary material) $\Delta H_{\text{NCO-CO}}$/kJ·mol$^{-1}$: $-20.0$ (2-butanone); $-16.2$ (4-heptanone). This means that amide-alkanone interactions are stronger in DMF solutions. The higher $H_{\mathrm{m}}^{\mathrm{E}}$ values of DMF systems can be explained assuming that the positive difference $\Delta H_{\text{NCO-NCO}}$ (DMF) − $\Delta H_{\text{NCO-NCO}}$ (NMP) is predominant over the negative $\Delta H_{\text{NCO-CO}}$ (DMF) − $\Delta H_{\text{NCO-CO}}$ (NMP) term.

A more exact treatment should take into the contribution from structural effects to $H_{\mathrm{m}}^{\mathrm{E}}$. For this purpose, the excess molar internal energy at constant volume, $U_{\mathrm{Vm}}^{\mathrm{E}}$, is an useful magnitude. It is calculated from [43, 55]:

$$U_{V\mathrm{m}}^{\mathrm{E}} = H_{\mathrm{m}}^{\mathrm{E}} - \frac{T\alpha_p V_{\mathrm{m}}^{\mathrm{E}}}{\kappa_T} \quad (19)$$

where $\frac{\alpha_p}{\kappa_T} T V_m^{\mathrm{E}}$ is the so-called equation of state (eos) contribution to $H_{\mathrm{m}}^{\mathrm{E}}$, and $\alpha_p$ and $\kappa_T$ stand for the isobaric thermal expansion coefficient and isothermal compressibility of the mixture, respectively. Along calculations, the needed $\kappa_T$ data were obtained from:

$$\kappa_T = \kappa_S + \frac{T\alpha_{\mathrm{p}}^2 V_{\mathrm{m}}}{C_{p,\mathrm{m}}} \quad (20)$$

assuming that $C_{p\mathrm{m}}^{\mathrm{E}}$ = 0,what is a good approximation in view of the low $H_{\mathrm{m}}^{\mathrm{E}}$ values of the investigated mixtures [62]. Thus, $U_{V\mathrm{m}}^{\mathrm{E}}$(DMF)/J·mol$^{-1}$ = 201 (acetone); 262 (2-butanone) and 303 (2-pentanone) [13], and $U_{V\mathrm{m}}^{\mathrm{E}}$(NMP)/J·mol$^{-1}$ = 26 (2-butanone); 351 (4-heptanone) (Table 8). The large differences between $U_{V\mathrm{m}}^{\mathrm{E}}$ and $H_{\mathrm{m}}^{\mathrm{E}}$ values must be remarked, as clearly show, in the present case, the importance of structural effects on $H_{\mathrm{m}}^{\mathrm{E}}$. Thus, the negative $H_{\mathrm{m}}^{\mathrm{E}}$ value of the NMP + 2-butanone mixture is due to the mentioned effects; it does not mean that interactions between unlike molecules are preponderant. Note that the positive $U_{V\mathrm{m}}^{\mathrm{E}}$ values obtained indicate that dominant contribution to this excess function arises from the breaking of interactions between like molecules.

### *6.2 Volumetric data*

It is well known that negative $V_{\mathrm{m}}^{\mathrm{E}}$ values reveal the existence of interactions between unlike molecules and/or structural effects (geometrical factors as differences in size and shape between them [63-65] or interstitial accommodation [64]).

Interestingly, the excess functions $U_{V\mathrm{m}}^{\mathrm{E}}$ and $V_{\mathrm{m}}^{\mathrm{E}}$ for tertiary amide + *n*-alkanone systems show opposite signs. This shows that the main contribution to $V_{\mathrm{m}}^{\mathrm{E}}$ comes from structural effects [63]. On the other hand, for a given tertiary amide, both $U_{V\mathrm{m}}^{\mathrm{E}}$ (or $H_{\mathrm{m}}^{\mathrm{E}}$) and $V_{\mathrm{m}}^{\mathrm{E}}$ change in line, as these magnitudes increase with *u*. In consequence, the observed $V_{\mathrm{m}}^{\mathrm{E}}$ variation can be attributed to changes in the interactional contribution to $V_{\mathrm{m}}^{\mathrm{E}}$. This means that the contributions which increase $V_{\mathrm{m}}^{\mathrm{E}}$, weakening of the amide-alkanone interactions and larger number of broken amide-amide interactions, are predominant over that arising from the disruption of ketone-ketone interactions, and suggests that the negative $V_{\mathrm{m}}^{\mathrm{E}}$ values may be also ascribed to the rather strong amide-ketone interactions formed along the mixing process. The symmetry of the $V_m^E$ curves may be discussed in terms of the difference in size between the mixture components. For DMA or NMP + 2-propanone systems, the $V_m^E$ curves are shifted to larger mole fractions of the ketone, the smaller component (this work, Figure 1, [15]). The same behaviour is observed when *u* increases in systems with a given tertiary amide as the $V_m^E$ curves become skewed towards increasing values of amide concentration (smaller compound). This is particularly remarkable for the DMF + 2-heptanone mixture [13]. However, others effects are also present. Both components are of similar size in the DMF + 2-propanone mixture, and the $V_m^E$ curve is also shifted to lower mole fractions of the amide [13]. This may be interpreted assuming that interactions between unlike molecules are more probable in that concentration range. For a given 2-alkanone, say 2-pentanone, $V_m^E$ /cm$^3$·mol$^{-1}$ changes in the order: $-0.243$ (DMA) > $-0.251$ (DMF) [13] > $-0.573$ (NMP) [15]. The lower $V_m^E$ value of the DMF system compared to that of the DMA solution may be due to interactions between unlike molecules are more important when DMF is involved, as the amide group is less sterically hindered. The $V_m^E$ value of the NMP mixture is noticeably lower than the corresponding value of the DMF solution and underlines the importance of structural effects in NMP systems.

Both $A_p = (\frac{\Delta V_m^E}{\Delta T})_p$ and $\alpha_p^E$ are negative for the studied systems (Table 4). Thus, $A_p$ (DMA)/cm$^3$·mol$^{-1}$·K$^{-1}$= $-4.6\cdot10^{-3}$; (acetone); $-2.7\cdot10^{-3}$ (2-butanone) and $-3.6\cdot10^{-3}$ (2-pentanone). These values are very similar to those encountered for DMF mixtures [13]. Negative $A_p$ and $\alpha_p^E$ values indicate that the structures are more difficult to be broken in the mixtures than in the pure liquids, which may be ascribed to the existence of interactions between unlike molecules and/or structural effects. For example, the 2-ethoxyethanol + octane, or 1-pentanol + cyclohexane systems, characterized by strong interactions between like molecules, show positive $A_p$ values over the entire mole fraction range ($7.6\cdot10^{-3}$ [66] and $2.3\cdot10^{-3}$ [67], respectively (values in cm$^3$·mol$^{-1}$·K$^{-1}$). Interactions between unlike molecules are dominant in the $CHCl_3$ + 1-butylamine mixture ($H_m^E$/J·mol$^{-1}$ values are $-3125$; $T$ = 303.15 K) [68]), and $A_p$ is negative, $-4.2\cdot10^{-3}$ cm$^3$·mol$^{-1}$·K$^{-1}$ [65]. This magnitude is also negative for the hexane + hexadecane system characterized by structural effects and $A_p$ is $-1.3\cdot10^{-2}$ (same units) for the hexane + hexadecane mixture [69].

Negative $\kappa_S^E$ values have been also explained in terms of structural effects and/or interactions between unlike molecules, while positive values are considered to be due to the breaking of physical interactions [70]. The negative $\kappa_S^E$ values of DMA + 2-alkanone systems (

−62.0 (2-propanone); −42.8 (2-butanone); −34.1 (2-pentanone) and −14.9 (2-heptanone)) fit well within the trends stated above. DMF +2-alkanone systems behave similarly [13]. The signs of the excess functions $V_{\mathrm{m}}^{\mathrm{E}}, \kappa_S^E, c^{\mathrm{E}}$ are consistent between them. Thus, for a given *N,N*-dialkylamide, $V_{\mathrm{m}}^{\mathrm{E}}, \kappa_S^E$ values are negative, while $c^{\mathrm{E}}$ values are positive. In addition, $V_{\mathrm{m}}^{\mathrm{E}}, \kappa_S^E$ increase and $c^{\mathrm{E}}$ decreases when *u* is increased.

### *5.3 Internal pressures*

We have also determined the internal pressures of the mixtures, $P_{\mathrm{int}}$ [71-73]:

$$P_{\mathrm{int}} = \frac{\alpha_P T}{\kappa_T} - p \tag{21}$$

The main contributions to $P_{\mathrm{int}}$ arise from dispersion forces and weak dipole-dipole interactions [73]. For pure the liquids under study, $P_{\mathrm{int},i}$ / MPa = 470.2 (NMP); 456.5 (DMF): 449.4 (DMA); 330.7 (acetone); 333.2 (2-butanone); 325.7 (2-pentanone) and 326.7 (2-heptanone). Some interesting information can be obtained from the comparison of these values with the corresponding results for the cohesive energy density ($D_{\mathrm{ce}}$), which is measure of the total molecular cohesion (per $\mathrm{cm}^3$) of the liquid [73]. The magnitude is defined as [73]:

$$D_{\mathrm{ce}} = \frac{\Delta H_{\mathrm{vap}} - RT}{V_m} \tag{22}$$

where $\Delta H_{\mathrm{vap}}$ is the molar enthalpy of vaporization at 298.15 K and $R$ the gas constant. Using values of $\Delta H_{\mathrm{vap}}$ from [74], the values $D_{\mathrm{ce},i}$/MPa obtained are: 573.8 (DMF); 521.8 (NMP); 512.8 (DMA); 389.3 (acetone); 358.5 (2-butanone); 334.3 (2-pentanone) and 326.9 (2-heptanone). From the $P_{\mathrm{int},i}$ and $D_{\mathrm{ce},i}$ values, we can state: (i) the existence of very strong dipolar interactions between amide or ketone molecules in the case of 2-propanone or 2-butanone; (ii) Interactions between amide molecules become weaker in the sequence: DMF > NMP > NMA, as is has been previously shown. However, it seems that dispersive interactions are stronger in NMP. (iii) Dipolar interactions become weaker when the chain length of the 2-alkanone increases. For the DMA systems, $P_{\mathrm{int}}$ / MPa = 384.9 (acetone); 383.1 (2-butanone); 371.8 (2-pentanone) and 370.9 (2-heptanone), which are very close to those obtained for DMF solutions [13]. On the other hand, $P_{\mathrm{int}}$ can be also calculated with the equation [72]:

$$P_{\mathrm{int}} = \frac{RT}{x_1 v_{f1} + x_2 v_{f2} + V_{\mathrm{m}}^{\mathrm{E}}} - p \tag{23}$$

where $v_{fi} (= RT / (p + P^{*}_{\text{int},i}))$ is the free volume of the component $i$ [72]. The $P_{\text{int}}$ / MPa values calculated for DMA solutions determined from equation (23) are: 405.3 (acetone); 398.0 (2-butanone); 392.2 (2-pentanone) and 379.0 (2-heptanone). The differences with the experimental values are 5.3%, 3.9%, 5.5% and 2.2%, respectively. In the case of mixtures containing DMF, these differences are: 3.8 %, 4.2%, 3.9 and 0.0%. This demonstrates, apart from the consistency of our data, that the van der Waals equation holds in rather large extent for the investigated solution as equation (23) is derived from this equation of state. That is, the present mixtures are characterized in large extent by dispersive interactions.

### 6.4 *Speeds of sound*

The parameter $\chi = \left( \frac{c}{c^{\text{id}}} \right)^2 - 1$ is widely used to estimate the non-ideality of a system. High $\chi$ values characterize mixtures which show strong deviations from the ideal behavior, as the methanol + 2-pyrrolidone mixture ( $\chi = 0.8$ [75]). For DMA systems: $\chi = 0.090$ (acetone); 0.062 (2-butanone); 0.050 (2-pentanone); 0.023 (2-heptanone). These solutions are close to the ideal behavior in terms of the speed of sound.

Molecular interactions in liquid mixtures can be examined by means of the Rao's constant, $R$ , [76] (also termed, molar sound velocity, $R = V_{\text{m}} c^{1/3}$ ). If there is no association or if the degree of association is independent of the concentration, the Rao's constant shows a linear dependence on the molar fractions of the components ( $R = x_1 R_1 + x_2 R_2$ ) [76-78]. Systems characterized by complex formation show deviations from this behaviour [79]. For DMA solutions, the Rao's constant changes linearly with $x_1$ (Figure S2, supplementary material), and this suggests that association/solvatation effects can be neglected. The same occur in mixtures containing DMF.

### 6.5 *Refractive indices*

It is well known that $n_{\text{D}}$ values of a liquid mixture can be used to calculate the molar refraction, $R_{\text{m}}$ , of the system from the Lorentz-Lorentz equation [80, 81]:

$$R_{\text{m}} = \frac{n_{\text{D}}^2 - 1}{n_{\text{D}}^2 + 2} V_{\text{m}} \tag{25}$$

This magnitude is closely related to the dispersion forces of the system under study, as $n_{\text{D}}$ at optical wavelengths is related to the mean polarizability [81]. For pure amides, the values (in $\text{cm}^3 \cdot \text{mol}^{-1}$) are: 26.8 (NMP) > 24.3 (DMA) > 19.9 (DMF), and dispersive interactions decrease in the same order. For 2-alkanones, $R_{\text{m}}$ /$\text{cm}^3 \cdot \text{mol}^{-1}$ = 16.1 (2-propanone) < 20.7 (2-butanone < 34.6 (2-heptanone). That is dispersive ketone-ketone interactions become more relevant when the 2-alkanone size is increased. Accordingly with these results: $R_{\mathrm{m}}$ also increases with $u$ in DMF or DMA systems (Figure S3, supplementary material). Thus, $R_{\mathrm{m}}$ (DMA)/cm$^3$·mol$^{-1}$ = 20.3 (2-propanone) < 22.5 (2-butanone) < 24.9 (2-pentanone) < 29.5 (2-heptanone), and $R_{\mathrm{m}}$ (DMF)/cm$^3$·mol$^{-1}$ = 18.1 (2-propanone) < 20.4 (2-butanone) < 22.6 (2-pentanone) < 27.4 (2-heptanone). We also note that dispersive interactions are more relevant in DMA solutions than in those including DMF. Interestingly, the magnitude $P_{\mathrm{int}}V_{\mathrm{m}}$ has been proposed as a measure of the London dispersion energy, independently of the existence of strong specific interactions [82]. For DMA + 2-alkanone mixtures, $P_{\mathrm{int}}V_{\mathrm{m}}$ increases linearly with $R_{\mathrm{m}}$ (at equimolar composition), as $P_{\mathrm{int}}V_{\mathrm{m}} = 7.31 + 1.22R_{\mathrm{m}}$; $(r = 0.997)$. In addition, $P_{\mathrm{int}}V_{\mathrm{m}}$ linearly decreases with $\overline{\mu}$ of 2-alkanones $P_{\mathrm{int}}V_{\mathrm{m}} = 68.57 - 30.56\overline{\mu}$; $(r = 0.986)$. DMF mixtures show the same behavior [13] and one can conclude that that dispersive interactions become stronger when $u$ is increased in the investigated systems.

*6.5 Results from the Flory model*

In spite of the few $H_{\mathrm{m}}^{\mathrm{E}}$ data available for the investigated systems, some interesting conclusions can be provided from the model application. (i) $X_{12}$ increases with the chain length of the 2-alkanone in systems with a given amide (Table 7). In this theory, $X_{12}$ is proportional to $\Delta\eta / v_s^*$, where $v_s^*$ stands for the reduction volume for segment and $\Delta\eta = \eta_{11} + \eta_{22} - 2\eta_{12}$. The $\eta_{ij}$ magnitudes are positive and characterize the energy of interaction for a pair of neighbouring sites. As $\eta_{11}$ remains constant, the mentioned $X_{12}$ increase may be ascribed to $\eta_{12}$ decreases more sharply than $\eta_{22}$ does when $u$ increases. The former is linked to a weakening of interactions between unlike molecules, the latter reflects a weakening of ketone-ketone interactions. (ii) Interestingly, the mean relative standard deviation for $H_{\mathrm{m}}^{\mathrm{E}}$ for DMF systems (0.052) is much lower than the value for NMP solutions (0.368). That is, it seems that orientational effects are much more relevant in the latter systems, while the random mixing hypothesis is held, in large extent, for mixtures including DMF. Accordingly, the concentration dependence of $X_{12}$ is much weaker for DMF systems than for NMP solutions (Figure S1). (iii) As a trend, the model yields calculated $V_{\mathrm{m}}^{\mathrm{E}}$ values larger than the experimental values (Table 8), which may be due to the interactional contribution to this excess function is overestimated. Nevertheless, the variation of $V_{\mathrm{m}}^{\mathrm{E}}$ with $u$ is correctly represented (Table 8). In fact, the model correctly correlates the $V_{\mathrm{m}}^{\mathrm{E}}$ data using $X_{12}$ values determined from such measurements. (iv) $V_{\mathrm{m}}^{\mathrm{E}}$ data can be better examined by means of the Prigogine-Flory-Patterson model (PFP) [83], where $V_{\mathrm{m}}^{\mathrm{E}}$ is the result of the sum of three contributions: an interactional contribution, a curvature term and the so-called $P$* term. The second one depends on $-(\overline{V}_1-\overline{V}_2)^2$ and is always negative. The latter depends on $(P_1^*-P_2^*)(\overline{V}_1-\overline{V}_2)$. For all the systems considered here, $P_1^*>P_2^*$ ; $\overline{V}_1<\overline{V}_2$ and the $P$* term is always negative. Results included in Table 8 remarks the importance of structural effects in the investigated systems.

*6.6 Comparison with other systems*

Some volumetric data are available in the literature on *N*-alkylamide + 2-alkanone systems. Thus, $V_m^E$ ($T$ = 303.15 K)/cm$^3$·mol$^{-1}$ = − 0.524 (*N*-methylformamide + 2-butanone) [84]; − 0.499 (*N*-methylformamide + 2-pentanone) [85]; − 0.434 (*N*-methylacetamide + 2-butanone) [86]. These values are lower than the corresponding results for systems with DMF or DMA and can be ascribed to the existence solvatation effects. Note that secondary amides are self-associated species and can form H-bonds with 2-alkanones [87].

## Conclusions

Data on $\rho$, $c$, $n_D$, at have been reported, at different temperatures, for the systems DMA + $CH_3CO(CH_2)_{u-1}CH_3$ ($u$ = 1,2,3,5). From these data, the excess functions $V_m^E, \kappa_S^E$ and $c^E$ have been determined. Negative $V_m^E$ values reveal the existence of structural effects and interactions between unlike molecules. From calorimetric data available in the literature, it is shown that the mentioned interactions are stronger in systems with DMF than in NMP solutions, and become weaker when $u$ increases in systems with a given amide. Structural effects largely contribute to $H_m^E$. They are more relevant in mixtures containing NMP. The application of the Flory's model shows that the random mixing hypothesis is valid in large extent for DMF solutions, while NMP systems are characterized by rather strong orientational effects. From the $P_{int}V_m$ and $R_m$ values, it is concluded that dispersive interactions become more relevant for the systems with longer 2-alkanones, or when DMF is replaced by DMA in mixtures with a given ketone.

## Acknowledgements

The authors gratefully acknowledge the financial support received from the Consejería de Educación y Cultura of Junta de Castilla y León, under Project BU034U16. A. Cobos and F. Hevia are grateful to Ministerio de Educación, Cultura y Deporte for the grants FPU15/05456 and FPU14/04104, respectively.

TABLE 1

Sample description

| Chemical | CAS number | Source | Purity[a] | Analysis method |
|---|---|---|---|---|
| *N,N*-dimethylacetamide | 127-19-5 | Sigma-Aldrich | ≥ 0.995 | GC[b] |
| Propanone | 67-64-1 | Sigma-Aldrich | ≥ 0.998 | HPLC[c] |
| 2-butanone | 78-93-3 | Fluka | ≥ 0.995 | GC[b] |
| 2-pentanone | 107-87-9 | Sigma-Aldrich | ≥ 0.98 | FCC[d] |
| 2-heptanone | 110-43-0 | Sigma-Aldrich | ≥ 0.99 | GC[b] |

[a]in mass fraction; [b]gas chromatography; [c]High-performance liquid chromatography [d]flash column chromatography

TABLE 2

Physical properties[a] of pure compounds at temperature *T* and pressure *p* = 0.1 MPa.

| Property/*T* | *N,N*-dimethylacetamide | 2-propanone | 2-butanone | 2-pentanone | 2-heptanone |
|---|---|---|---|---|---|
| $\rho$/g cm$^{-3}$ | | | | | |
| *T*/K=293.15 | 0.94101 | 0.79018 | 0.80512 | 0.80654 | 0.81557 |
| | 0.9410[88] | 0.790546[89] | 0.80495[90] | 0.806322[91] | 0.81537[40] |
| | | 0.78998[40] | 0.8049[40] | 0.8064[40] | |
| | | | 0.805058[92] | 0.80626[90] | |
| *T*/K=298.15 | 0.93639 | 0.78441 | 0.79977 | 0.80166 | 0.81119 |
| | 0.93634[93] | 0.784431[54] | 0.79974[90] | 0.801522[91] | 0.81123[40] |
| | | 0.7844[40] | 0.7997[40] | 0.8015[40] | 0.81093[94] |
| | | | | 0.80142[90] | |
| *T*/K=303.15 | 0.93173 | 0.77876 | 0.79465 | 0.79688 | 0.80698 |
| | 0.93166[93] | 0.77966[90] | 0.79448[90] | 0.796723[91] | 0.806827[24] |
| | 0.93162[95] | 0.77863[13] | 0.7946[40] | 0.79658[90] | |
| | | | 0.794565[92] | | |
| *c*/m s$^{-1}$ | | | | | |
| *T*/K=293.15 | 1475.3 | 1182.0 | 1212.1 | 1232.5 | 1282.3 |
| | 1478.98[93] | 1182.5[89] | 1212.3[96] | 1231.65[91] | 1281.92[24] |
| | | 1185[90] | 1213[90] | 1233[90] | |
| *T*/K=298.15 | 1456.0 | 1164.2 | 1190.8 | 1211.8 | 1263.0 |
| | 1455.91[93] | 1162.0[96] | 1191.6[96] | 1211.17[91] | 1262.49[24] |
| | 1456.42[95] | 1161.7[54] | 1192[90] | 1213[90] | |
| | 1458[97] | 1163.9[98] | | | |
| *T*/K=303.15 | 1435.9 | 1139.3 | 1170.9 | 1192.3 | 1244.7 |
| | 1435.55[93] | 1139.2[89] | 1171[90] | 1191.79[91] | 1244.14[24] |
| | | 1140[90] | | 1192[90] | |
| $\alpha_p$/10$^{-3}$K$^{-1}$ | | | | | |
| *T*/K=298.15 | 0.98 | 1.46 | 1.31 | 1.2 | 1.06 |
| | 0.960[99] | 1.45[89] | 1.31[90] | 1.22[40] | 1.06[40] |
| | | 1.46[92] | | 1.21[92] | |

TABLE 2 (continued)

| | | | | | |
|---|---|---|---|---|---|
| $\kappa_S$/TPa$^{-1}$ | | | | | |
| $T$/K=293.15 | 488.34 | 905.81 | 845.40 | 816.21 | 745.69 |
| | | 904.66[89] | 844[90] | 816[90] | 746.26[24] |
| | | 900[92] | | | |
| $T$/K=298.15 | 503.68 | 940.64 | 881.77 | 849.47 | 772.76 |
| | 503.85[93] | 946[100] | 880[90] | 849[90] | 773.48[24] |
| | | 944.6[50] | | 848[92] | |
| | | 941.1[98] | | | |
| $T$/K=303.15 | 520.25 | 989.28 | 917.88 | 882.75 | 799.91 |
| | 516[101] | 988.80[89] | 917.9[102] | 884[90] | 800.71[24] |
| | | | 918[90] | | |
| $\kappa_T$/TPa$^{-1}$ | | | | | |
| $T$/K=298.15 | 656.7 | 1315.7 | 1171.9 | 1098.3 | 967.2 |
| | 671[103] | 1324[40] | 1175.9[102] | 1092[40] | 957[40] |
| | | | 1188[36] | | |
| $C_{p,\mathrm{m}}$/J mol$^{-1}$ K$^{-1}$ | | | | | |
| $T$/K=298.15 | 175.5[104] | 125.45[90] | 159[105] | 185.4[106] | 242.54[94] |
| $n_\mathrm{D}$ | | | | | |
| $T$/K=293.15 | 1.43845 | 1.35854 | 1.37839 | 1.39016 | |
| | 1.4384[40] | 1.35868[40] | 1.3788[40] | 1.39080[40] | |
| | | 1.3584[107] | | | |
| $T$/K=298.15 | 1.43621 | 1.35386 | 1.37612 | 1.38773 | 1.40688 |
| | 1.4357][108] | 1.35597[109] | 1.3764[110] | 1.3885[111] | 1.40655[40] |
| | 1.4359[112] | | 1.3767[111] | | |
| | 1.4364[113] | | | | |
| $T$/K=303.15 | 1.43367 | 1.35323 | 1.37351 | 1.38530 | |
| | 1.4342[113] | | 1.3740[114] | | |

[a]$\rho$, density; $c$, speed of sound; $\alpha_p$, isobaric thermal expansion coefficient; $\kappa_S$, adiabatic compressibility; $\kappa_T$, isothermal compressibility; $C_{p,\mathrm{m}}$, isobaric molar heat capacity and $n_\mathrm{D}$, refractive index of pure components. Relative standard uncertainties, $u_r$, are: $u_r(\rho^*) = 0.0012$; $u_r(c^*) = 0.0004$; $u_r(\alpha_p) = 0.028$; $u_r(\kappa_S^*) = 0.002$; $u_r(\kappa_T^*) = 0.015$; $u_r(n_D^*) = 0.0015$; standard uncertainties for temperature and pressure are $u(T) = 0.02\mathrm{K}$ $u(T) = 0.02\mathrm{K}$ (for $n_\mathrm{D}$ values, $u(T) = 0.02\mathrm{K}$); and $u(\rho) = 1\mathrm{kPa}$.

TABLE 3

Densities, $\rho$, speeds of sound, $c$, adiabatic compressibility, $\kappa_S$, and molar excess volumes, $V_m^E$, for *N,N*-dimethylacetamide (1) + 2-alkanone (2) mixtures at temperature $T$ and 0.1 MPa.

| $x_1$ | $\rho$ / gcm$^{-3}$ | $c$ / ms$^{-1}$ | $\kappa_S$ / TPa$^{-1}$ | $V_m^E$ / cm$^3$mol$^{-1}$ | $x_1$ | $\rho$ / gcm$^{-3}$ | $c$ / ms$^{-1}$ | $\kappa_S$ / TPa$^{-1}$ | $V_m^E$ / cm$^3$mol$^{-1}$ |
|---|---|---|---|---|---|---|---|---|---|
| | | | | *N,N*-dimethylacetamide (1) + 2-propanone (2) $T$/K= 293.15 | | | | | |
| 0.0000 | 0.79018 | 1182.0 | 905.81 | 0.0000 | 0.5524 | 0.88569 | 1358.0 | 612.24 | − 0.3588 |
| 0.0558 | 0.80177 | 1201.8 | 863.55 | − 0.1066 | 0.5974 | 0.89201 | 1370.9 | 596.51 | − 0.3458 |
| 0.1019 | 0.81090 | 1217.7 | 831.67 | − 0.1740 | 0.6502 | 0.89910 | 1385.2 | 579.65 | − 0.3151 |
| 0.1470 | 0.81947 | 1232.9 | 802.81 | − 0.2239 | 0.7022 | 0.90595 | 1399.4 | 563.65 | − 0.2881 |
| 0.2021 | 0.82962 | 1251.1 | 770.08 | − 0.2778 | 0.7502 | 0.91209 | 1412.3 | 549.68 | − 0.2592 |
| 0.2523 | 0.83856 | 1267.4 | 742.40 | − 0.3194 | 0.8004 | 0.91813 | 1425.0 | 536.37 | − 0.2057 |
| 0.2991 | 0.84651 | 1282.2 | 718.55 | − 0.3398 | 0.8538 | 0.92453 | 1438.6 | 522.64 | − 0.1609 |
| 0.3525 | 0.85539 | 1298.9 | 692.92 | − 0.3650 | 0.8980 | 0.92965 | 1449.7 | 511.83 | − 0.1177 |
| 0.3991 | 0.86282 | 1313.2 | 672.08 | − 0.3739 | 0.9468 | 0.93513 | 1461.6 | 500.58 | − 0.0643 |
| 0.4452 | 0.86994 | 1327.1 | 652.68 | − 0.3759 | 1.0000 | 0.94091 | 1474.2 | 489.03 | 0.0000 |
| 0.4994 | 0.87804 | 1342.8 | 631.63 | − 0.3709 | | | | | |
| | | | | *N,N*-dimethylacetamide (1) +2-propanone (2) $T$/K= 298.15 | | | | | |
| 0.0000 | 0.78441 | 1164.2 | 940.64 | 0.0000 | 0.5547 | 0.88094 | 1343.9 | 628.55 | − 0.3716 |
| 0.0549 | 0.79597 | 1184.2 | 895.93 | | 0.6113 | 0.88889 | 1360.2 | 608.10 | − 0.3519 |
| 0.1023 | 0.80533 | 1200.5 | 861.54 | − 0.1789 | 0.6531 | 0.89464 | 1372.1 | 593.73 | − 0.3378 |
| 0.1573 | 0.81599 | 1219.6 | 823.94 | − 0.2573 | 0.7026 | 0.90105 | 1385.4 | 578.21 | − 0.2963 |
| 0.2021 | 0.82421 | 1234.6 | 795.95 | − 0.2964 | 0.7528 | 0.90760 | 1399.3 | 562.68 | |
| 0.2522 | 0.83319 | 1251.2 | 766.65 | − 0.3387 | 0.7985 | 0.91314 | 1411.1 | 549.99 | − 0.2225 |
| 0.2989 | 0.84118 | 1266.2 | 741.47 | − 0.3595 | 0.8470 | 0.91890 | 1423.5 | 537.09 | − 0.1699 |
| 0.3512 | 0.84993 | 1283.0 | 714.74 | − 0.3842 | 0.9009 | 0.92523 | 1437.2 | 523.25 | − 0.1179 |
| 0.3995 | 0.85775 | 1298.0 | 692.00 | − 0.3991 | 0.9441 | 0.93022 | 1448.2 | 512.56 | − 0.0790 |
| 0.4457 | 0.86490 | 1312.1 | 671.63 | − 0.3977 | 1.0000 | 0.93625 | 1461.5 | 500.05 | 0.0000 |

TABLE 3 (continued)

| | | | | | | | | | |
|---|---|---|---|---|---|---|---|---|---|
| 0.4963 | 0.87248 | 1327.0 | 650.88 | − 0.3890 | | | | | |

*N,N*-dimethylacetamide (1) + 2-propanone (2) *T*/K= 303.15

| | | | | | | | | | |
|---|---|---|---|---|---|---|---|---|---|
| 0.0000 | 0.77876 | 1139.3 | 989.28 | 0.0000 | 0.5524 | 0.87586 | 1318.9 | 656.36 | − 0.4060 |
| 0.0558 | 0.79051 | 1159.5 | 940.92 | − 0.1146 | 0.5974 | 0.88221 | 1331.7 | 639.17 | − 0.3854 |
| 0.1019 | 0.79982 | 1175.6 | 904.67 | − 0.1922 | 0.6502 | 0.88949 | 1346.4 | 620.17 | − 0.3605 |
| 0.1470 | 0.80859 | 1191.3 | 871.42 | − 0.2536 | 0.7022 | 0.89635 | 1360.5 | 602.73 | − 0.3220 |
| 0.2021 | 0.81888 | 1209.9 | 834.22 | − 0.3112 | 0.7502 | 0.90252 | 1373.7 | 587.16 | − 0.2844 |
| 0.2523 | 0.82795 | 1226.8 | 802.51 | − 0.3559 | 0.8004 | 0.90870 | 1386.4 | 572.54 | − 0.2319 |
| 0.2991 | 0.83609 | 1241.7 | 775.74 | − 0.3851 | 0.8538 | 0.91523 | 1400.3 | 557.22 | − 0.1866 |
| 0.3525 | 0.84512 | 1258.7 | 746.86 | − 0.4140 | 0.8980 | 0.92050 | 1411.7 | 545.12 | − 0.1470 |
| 0.3991 | 0.85263 | 1273.1 | 723.63 | − 0.4206 | 0.9468 | 0.92597 | 1423.5 | 532.95 | − 0.0801 |
| 0.4452 | 0.85985 | 1287.2 | 701.92 | − 0.4222 | 1.0000 | 0.93173 | 1436.0 | 520.48 | 0.0000 |
| 0.4994 | 0.86808 | 1303.4 | 678.09 | − 0.4176 | | | | | |

*N,N*-dimethylacetamide (1) + 2-butanone (2) *T*/K= 293.15

| | | | | | | | | | |
|---|---|---|---|---|---|---|---|---|---|
| 0.0000 | 0.80512 | 1212.1 | 845.40 | 0.0000 | 0.5510 | 0.88334 | 1358.1 | 613.77 | − 0.2290 |
| 0.0585 | 0.81389 | 1227.8 | 815.04 | − 0.0625 | 0.5993 | 0.88979 | 1370.6 | 598.26 | − 0.2195 |
| 0.1098 | 0.82153 | 1241.6 | 789.61 | | 0.6556 | 0.89724 | 1385.4 | 580.69 | − 0.2043 |
| 0.1550 | 0.82802 | 1253.3 | 768.86 | − 0.1343 | 0.7049 | 0.90371 | 1398.3 | 565.94 | − 0.1880 |
| 0.2085 | 0.83580 | 1267.5 | 744.73 | − 0.1715 | 0.7570 | 0.91041 | 1411.9 | 551.00 | − 0.1596 |
| 0.2537 | 0.84222 | 1279.4 | 725.37 | − 0.1884 | 0.8008 | 0.91605 | 1423.4 | 538.80 | − 0.1385 |
| 0.3022 | 0.84912 | 1292.3 | 705.19 | − 0.2098 | 0.8488 | 0.92223 | 1436.1 | 525.76 | − 0.1175 |
| 0.3501 | 0.85581 | 1305.3 | 685.81 | − 0.2197 | 0.9052 | 0.92940 | 1450.9 | 511.12 | − 0.0866 |
| 0.3997 | 0.86270 | 1318.0 | 667.28 | − 0.2282 | 0.9469 | 0.93449 | 1461.5 | 500.99 | − 0.0448 |
| 0.4566 | 0.87053 | 1333.1 | 646.38 | − 0.2330 | 1.0000 | 0.94103 | 1475.3 | 488.24 | 0.0000 |
| 0.5019 | 0.87670 | 1344.9 | 630.62 | − 0.2324 | | | | | |

TABLE 3 (continued)

| | | | | | | | | | |
|---|---|---|---|---|---|---|---|---|---|
| *N,N*-dimethylacetamide (1) + 2-butanone (2) $T$/K= 298.15 | | | | | | | | | |
| 0.0000 | 0.79977 | 1190.8 | 881.77 | 0.0000 | 0.5412 | 0.87713 | 1335.1 | 639.60 | − 0.2450 |
| 0.0551 | 0.80810 | 1205.7 | 851.25 | − 0.0641 | 0.6016 | 0.88529 | 1351.1 | 618.78 | − 0.2376 |
| 0.1073 | 0.81591 | 1219.8 | 823.72 | − 0.1176 | 0.6560 | 0.89249 | 1365.4 | 601.00 | − 0.2184 |
| 0.1536 | 0.82269 | 1232.1 | 800.70 | − 0.1505 | 0.7067 | 0.89915 | 1378.8 | 585.01 | − 0.1979 |
| 0.2041 | 0.83005 | 1245.7 | 776.37 | − 0.1843 | 0.7524 | 0.90509 | 1390.8 | 571.19 | − 0.1751 |
| 0.2536 | 0.83718 | 1258.8 | 753.82 | − 0.2103 | 0.8052 | 0.91194 | 1404.8 | 555.65 | − 0.1500 |
| 0.2936 | 0.84287 | 1269.4 | 736.28 | − 0.2250 | 0.8521 | 0.91791 | 1417.0 | 542.58 | − 0.1183 |
| 0.3473 | 0.85044 | 1283.6 | 713.67 | − 0.2395 | 0.9056 | 0.92466 | 1431.0 | 528.13 | − 0.0788 |
| 0.4052 | 0.85850 | 1299.0 | 690.31 | − 0.2471 | 0.9459 | 0.92971 | 1441.5 | 517.63 | − 0.0475 |
| 0.4522 | 0.86500 | 1311.5 | 672.12 | − 0.2510 | 1.0000 | 0.93641 | 1455.6 | 504.02 | 0.0000 |
| 0.4958 | 0.87095 | 1323.0 | 655.97 | − 0.2479 | | | | | |
| *N,N*-dimethylacetamide (1) + 2-butanone (2) $T$/K= 303.15 | | | | | | | | | |
| 0.0000 | 0.79465 | 1170.9 | 917.88 | 0.0000 | 0.5418 | 0.87242 | 1330.1 | 647.90 | − 0.2543 |
| 0.0549 | 0.80298 | 1189.2 | 880.61 | − 0.0657 | 0.6085 | 0.88143 | 1347.5 | 624.82 | − 0.2418 |
| 0.1048 | 0.81039 | 1205.2 | 849.55 | − 0.1089 | 0.6488 | 0.88677 | 1356.1 | 613.20 | − 0.2253 |
| 0.1512 | 0.81725 | 1219.9 | 822.24 | − 0.1472 | 0.7015 | 0.89370 | 1369.5 | 596.60 | − 0.2006 |
| 0.2033 | 0.82482 | 1235.7 | 793.99 | − 0.1773 | 0.7563 | 0.90092 | 1382.7 | 580.57 | − 0.1792 |
| 0.2506 | 0.83166 | 1249.6 | 770.04 | − 0.2029 | 0.7981 | 0.90631 | 1392.1 | 569.35 | − 0.1526 |
| 0.3050 | 0.83950 | 1265.6 | 743.68 | − 0.2315 | 0.8467 | 0.91259 | 1402.5 | 557.08 | − 0.1252 |
| 0.3451 | 0.84521 | 1276.9 | 725.64 | − 0.2465 | 0.8954 | 0.91874 | 1414.2 | 544.23 | − 0.0855 |
| 0.3958 | 0.85231 | 1291.4 | 703.53 | − 0.2546 | 0.9441 | 0.92489 | 1423.6 | 533.50 | − 0.0481 |
| 0.4492 | 0.85976 | 1306.2 | 681.72 | − 0.2624 | 1.0000 | 0.93187 | 1436.2 | 520.25 | 0.0000 |
| 0.4871 | 0.86497 | 1315.7 | 667.86 | − 0.2612 | | | | | |

TABLE 3 (continued)

*N,N*-dimethylacetamide (1) + 2-pentanone (2) $T$/K= 293.15

| | | | | | | | | | |
|---|---|---|---|---|---|---|---|---|---|
| 0.0000 | 0.80654 | 1232.5 | 816.21 | 0.0000 | 0.5550 | 0.87800 | 1356.7 | 618.78 | – 0.1797 |
| 0.0530 | 0.81307 | 1243.3 | 795.65 | – 0.0402 | 0.6075 | 0.88514 | 1369.6 | 602.28 | – 0.1713 |
| 0.1061 | 0.81966 | 1254.5 | 775.22 | – 0.0735 | 0.6513 | 0.89116 | 1380.8 | 588.55 | – 0.1624 |
| 0.1551 | 0.82579 | 1264.9 | 756.86 | – 0.0997 | 0.6982 | 0.89763 | 1393.2 | 573.95 | |
| 0.2004 | 0.83149 | 1274.7 | 740.16 | – 0.1190 | 0.7407 | 0.90361 | 1403.8 | 561.58 | – 0.1365 |
| 0.2552 | 0.83848 | 1287.0 | 720.03 | – 0.1401 | 0.7992 | 0.91192 | 1419.5 | 544.22 | – 0.1178 |
| 0.3070 | 0.84515 | 1298.3 | 701.97 | – 0.1567 | 0.8532 | 0.91960 | 1434.0 | 528.81 | – 0.0884 |
| 0.3637 | 0.85253 | 1311.2 | 682.26 | – 0.1706 | 0.9004 | 0.92645 | 1447.1 | 515.44 | – 0.0664 |
| 0.4083 | 0.85837 | 1321.5 | 667.10 | – 0.1763 | 0.9439 | 0.93271 | 1459.1 | 503.60 | – 0.0319 |
| 0.4566 | 0.86477 | 1333.0 | 650.79 | – 0.1802 | 1.0000 | 0.94101 | 1475.2 | 488.32 | 0.0000 |
| 0.5013 | 0.87074 | 1343.6 | 636.17 | – 0.1813 | | | | | |

*N,N*-dimethylacetamide (1) + 2-pentanone (2) $T$/K= 298.15

| | | | | | | | | | |
|---|---|---|---|---|---|---|---|---|---|
| 0.0000 | 0.80166 | 1211.8 | 849.47 | 0.0000 | 0.5497 | 0.87265 | 1335.5 | 642.50 | – 0.2002 |
| 0.0540 | 0.80832 | 1223.0 | 827.11 | – 0.0414 | 0.6041 | 0.88007 | 1348.9 | 624.49 | – 0.1930 |
| 0.1037 | 0.81457 | 1233.6 | 806.72 | – 0.0834 | 0.6373 | 0.88465 | 1357.5 | 613.41 | – 0.1879 |
| 0.1497 | 0.82034 | 1243.5 | 788.34 | – 0.1100 | 0.6983 | 0.89309 | 1373.1 | 593.88 | – 0.1686 |
| 0.2016 | 0.82693 | 1254.7 | 768.16 | – 0.1386 | 0.7431 | 0.89936 | 1384.9 | 579.74 | – 0.1519 |
| 0.2553 | 0.83381 | 1266.3 | 747.93 | – 0.1629 | 0.7979 | 0.90714 | 1399.3 | 562.99 | – 0.1312 |
| 0.2990 | 0.83945 | 1276.3 | 731.31 | – 0.1781 | 0.8446 | 0.91383 | 1412.2 | 548.71 | – 0.1095 |
| 0.3427 | 0.84513 | 1286.4 | 715.03 | – 0.1895 | 0.8957 | 0.92124 | 1426.3 | 533.59 | – 0.0842 |
| 0.3951 | 0.85199 | 1298.6 | 696.01 | – 0.1978 | 0.9392 | 0.92753 | 1438.4 | 521.09 | – 0.0518 |
| 0.4452 | 0.85862 | 1310.3 | 678.36 | – 0.2027 | 1.0000 | 0.93639 | 1455.5 | 504.10 | 0.0000 |
| 0.4986 | 0.86576 | 1323.0 | 659.91 | – 0.2041 | | | | | |

TABLE 3 (continued)

| | | | | | | | | | |
|---|---|---|---|---|---|---|---|---|---|
| *N,N*-dimethylacetamide (1) + 2-pentanone (2) *T*/K= 303.15 | | | | | | | | | |
| 0.0000 | 0.79688 | 1192.3 | 882.75 | 0.0000 | 0.5495 | 0.86798 | 1316.5 | 664.74 | − 0.2150 |
| 0.0535 | 0.80355 | 1203.6 | 859.06 | − 0.0512 | 0.5945 | 0.87410 | 1327.4 | 649.28 | − 0.2070 |
| 0.0989 | 0.80923 | 1213.4 | 839.31 | − 0.0865 | 0.6408 | 0.88047 | 1339.2 | 633.28 | − 0.1976 |
| 0.1528 | 0.81603 | 1225.0 | 816.62 | − 0.1232 | 0.6965 | 0.88823 | 1353.4 | 614.64 | − 0.1843 |
| 0.2015 | 0.82226 | 1235.4 | 796.85 | − 0.1556 | 0.7468 | 0.89529 | 1366.7 | 597.98 | − 0.1661 |
| 0.2532 | 0.82887 | 1247.2 | 775.61 | − 0.1773 | 0.7993 | 0.90273 | 1380.4 | 581.34 | − 0.1427 |
| 0.3051 | 0.83556 | 1258.7 | 755.40 | − 0.1935 | 0.8450 | 0.90926 | 1393.0 | 566.77 | − 0.1181 |
| 0.3498 | 0.84140 | 1269.2 | 737.80 | − 0.2072 | 0.9015 | 0.91740 | 1408.4 | 549.53 | − 0.0820 |
| 0.3969 | 0.84757 | 1280.2 | 719.89 | − 0.2136 | 0.9476 | 0.92407 | 1421.3 | 535.70 | − 0.0451 |
| 0.4491 | 0.85446 | 1292.2 | 700.89 | − 0.2151 | 1.0000 | 0.93173 | 1435.9 | 520.55 | 0.0000 |
| 0.4998 | 0.86125 | 1304.3 | 682.52 | − 0.2164 | | | | | |
| *N,N*-dimethylacetamide (1) + 2-heptanone (2) *T*/K= 298.15 | | | | | | | | | |
| 0.0000 | 0.81119 | 1263.0 | 772.76 | 0.0000 | 0.5571 | 0.86810 | 1345.0 | 636.81 | − 0.0051 |
| 0.0578 | 0.81613 | 1270.0 | 759.67 | − 0.0087 | 0.6080 | 0.87466 | 1354.9 | 622.76 | − 0.0067 |
| 0.1069 | 0.82044 | 1276.0 | 748.62 | − 0.0109 | 0.6492 | 0.88020 | 1363.6 | 611.03 | − 0.0107 |
| 0.1570 | 0.82502 | 1282.3 | 737.10 | − 0.0161 | 0.6981 | 0.88699 | 1374.2 | 597.00 | − 0.0092 |
| 0.2088 | 0.82984 | 1289.1 | 725.13 | − 0.0071 | 0.7499 | 0.89455 | 1386.2 | 581.78 | − 0.0115 |
| 0.2529 | 0.83418 | 1295.1 | 714.67 | − 0.0126 | 0.8011 | 0.90234 | 1398.8 | 566.42 | − 0.0092 |
| 0.3080 | 0.83970 | 1303.0 | 701.38 | − 0.0052 | 0.8520 | 0.91047 | 1412.1 | 550.79 | − 0.0082 |
| 0.3485 | 0.84399 | 1309.3 | 691.17 | − 0.0114 | 0.9015 | 0.91879 | 1426.0 | 535.22 | − 0.0104 |
| 0.4006 | 0.84961 | 1317.4 | 678.14 | − 0.0050 | 0.9516 | 0.92755 | 1440.7 | 519.40 | − 0.0035 |
| 0.4484 | 0.85472 | 1324.8 | 666.58 | | 1.0000 | 0.93646 | 1456.0 | 503.68 | 0.0000 |
| 0.5048 | 0.86163 | 1335.1 | 651.09 | | | | | | |

Uncertainties are: $u(x_1) = 0.0001$; $u(p) = 1\text{kPa}$; $u(T) = 0.01\text{K}$; and the combined expanded uncertainties (0.95 level of confidence) are: $U_{rc}(\rho) = 0.020\text{kgm}^{-3}$; $U_{rc}(c) = 0.2\text{ms}^{-1}$; $U_{rc}(\kappa_S) = 0.02\text{TPa}^{-1}$; $U_{rc}(V_m^{\ E}) = (0.02\,|\,V_{\text{m,max}}^{\ \text{E}}\,|+0.005)\,\text{cm}^3\ \text{mol}^{-1}$.

TABLE 4

Isobaric thermal expansion coefficient, $\alpha_p$, and excess functions, at 298.15 K and 0.1 MPa, for adiabatic compressibility, $\kappa_S$, speed of sound, *c*, and isobaric thermal expansion coefficient, $\alpha_p^{\mathrm{E}}$, of *N,N*-dimethylacetamide (1) + 2-alkanone (2) mixtures.

| $x_1$ | $\alpha_p$ / $10^{-3}$ K$^{-1}$ | $\kappa_S^{\mathrm{E}}$ / TPa$^{-1}$ | $c^{\mathrm{E}}$ / ms$^{-1}$ | $\alpha_p^{\mathrm{E}}$ / $10^{-3}$ K$^{-1}$ | $x_1$ | $\alpha_p$ / $10^{-3}$ K$^{-1}$ | $\kappa_S^{\mathrm{E}}$ / TPa$^{-1}$ | $c^{\mathrm{E}}$ / ms$^{-1}$ | $\alpha_p^{\mathrm{E}}$ / $10^{-3}$ K$^{-1}$ |
|---|---|---|---|---|---|---|---|---|---|
| *N,N*-dimethylacetamide (1) + 2-propanone (2) *T*/K= 298.15 | | | | | | | | | |
| 0.0000 | 1.46 | 0.0 | 0.0 | 0.000 | 0.5547 | 1.12 | − 58.7 | 55.8 | − 0.050 |
| 0.0549 | 1.41 | − 19.1 | 11.6 | − 0.019 | 0.6113 | 1.09 | − 54.4 | 54.4 | − 0.047 |
| 0.1023 | 1.37 | − 31.4 | 19.9 | − 0.029 | 0.6531 | 1.08 | − 50.6 | 52.4 | − 0.044 |
| 0.1573 | 1.33 | − 43.5 | 29.0 | − 0.037 | 0.7026 | 1.06 | − 44.8 | 48.5 | − 0.041 |
| 0.2021 | 1.30 | − 50.7 | 35.2 | − 0.043 | 0.7528 | 1.04 | − 39.1 | 44.1 | − 0.036 |
| 0.2522 | 1.27 | − 56.9 | 41.4 | − 0.047 | 0.7985 | 1.03 | − 32.5 | 38.3 | − 0.031 |
| 0.2989 | 1.24 | − 60.6 | 46.0 | − 0.050 | 0.8470 | 1.01 | − 25.3 | 31.1 | − 0.026 |
| 0.3512 | 1.21 | − 63.5 | 50.5 | − 0.052 | 0.9009 | 1.00 | − 16.9 | 21.7 | − 0.018 |
| 0.3995 | 1.19 | − 64.4 | 53.4 | − 0.052 | 0.9441 | 0.99 | − 10.0 | 12.0 | − 0.011 |
| 0.4457 | 1.17 | − 64.0 | 55.4 | − 0.052 | 1.0000 | 0.98 | 0.0 | 0.0 | 0.000 |
| 0.4963 | 1.14 | − 62.1 | 56.2 | − 0.052 | | | | | |
| *N,N*-dimethylacetamide (1) + 2-butanone (2) *T*/K= 298.15 | | | | | | | | | |
| 0.0000 | 1.31 | 0.0 | 0.0 | 0.000 | 0.5412 | 1.10 | − 42.0 | 40.1 | − 0.027 |
| 0.0551 | 1.29 | − 10.7 | 7.1 | − 0.004 | 0.6016 | 1.08 | − 39.8 | 39.8 | − 0.024 |
| 0.1073 | 1.27 | − 19.4 | 13.3 | − 0.007 | 0.6560 | 1.07 | − 36.7 | 38.3 | − 0.021 |
| 0.1536 | 1.25 | − 25.5 | 18.1 | − 0.011 | 0.7067 | 1.06 | − 33.2 | 36.1 | − 0.017 |
| 0.2041 | 1.23 | − 31.3 | 23.2 | − 0.015 | 0.7524 | 1.05 | − 29.4 | 33.2 | − 0.013 |
| 0.2536 | 1.20 | − 35.6 | 27.3 | − 0.019 | 0.8052 | 1.03 | − 24.5 | 28.8 | − 0.009 |
| 0.2936 | 1.19 | − 38.4 | 30.3 | − 0.022 | 0.8521 | 1.02 | − 19.3 | 23.7 | − 0.005 |
| 0.3473 | 1.17 | − 41.0 | 33.7 | − 0.026 | 0.9056 | 1.01 | − 12.9 | 16.6 | − 0.002 |
| 0.4052 | 1.15 | − 42.7 | 36.7 | − 0.028 | 0.9459 | 1.00 | − 7.6 | 9.9 | − 0.001 |
| 0.4522 | 1.13 | − 43.2 | 38.4 | − 0.028 | 1.0000 | 0.98 | 0.0 | 0.0 | 0.000 |
| 0.4958 | 1.12 | − 42.9 | 39.5 | − 0.028 | | | | | |

TABLE 4 (continued)

| | | | | | | | | | |
|---|---|---|---|---|---|---|---|---|---|
| | | | *N,N*-dimethylacetamide (1) + 2-pentanone (2) *T*/K= 298.15 | | | | | | |
| 0.0000 | 1.20 | 0.0 | 0.0 | 0.000 | 0.5497 | 1.06 | −33.8 | 32.5 | −0.030 |
| 0.0540 | 1.18 | −6.9 | 4.9 | −0.008 | 0.6041 | 1.05 | −32.6 | 32.6 | −0.030 |
| 0.1037 | 1.16 | −12.8 | 9.2 | −0.016 | 0.6373 | 1.04 | −31.8 | 32.6 | −0.030 |
| 0.1497 | 1.15 | −17.5 | 13.0 | −0.022 | 0.6983 | 1.03 | −29.0 | 31.2 | −0.029 |
| 0.2016 | 1.14 | −22.0 | 16.8 | −0.026 | 0.7431 | 1.02 | −26.4 | 29.4 | −0.028 |
| 0.2553 | 1.12 | −25.6 | 20.2 | −0.029 | 0.7979 | 1.01 | −22.2 | 25.9 | −0.026 |
| 0.2990 | 1.11 | −28.5 | 23.1 | −0.030 | 0.8446 | 1.00 | −18.4 | 22.2 | −0.023 |
| 0.3427 | 1.10 | −30.8 | 25.7 | −0.030 | 0.8957 | 1.00 | −13.2 | 16.7 | −0.018 |
| 0.3951 | 1.09 | −32.8 | 28.3 | −0.030 | 0.9392 | 0.99 | −8.1 | 11.6 | −0.012 |
| 0.4452 | 1.08 | −33.8 | 30.1 | −0.030 | 1.0000 | 0.99 | 0.0 | 0.0 | 0.000 |
| 0.4986 | 1.07 | −34.1 | 31.6 | −0.030 | | | | | |
| | | | *N,N*-dimethylacetamide (1) + 2-heptanone (2) *T*/K= 298.15 | | | | | | |
| 0.0000 | 1.06 | 0.0 | 0.0 | | 0.5571 | | −15.3 | 15.8 | |
| 0.0578 | | −2.9 | 2.3 | | 0.6080 | | −15.3 | 16.2 | |
| 0.1069 | | −4.8 | 4.1 | | 0.6492 | | −15.1 | 16.5 | |
| 0.1570 | | −6.8 | 5.7 | | 0.6981 | | −14.5 | 16.3 | |
| 0.2088 | | −8.4 | 7.4 | | 0.7499 | | −13.4 | 15.7 | |
| 0.2529 | | −9.8 | 8.7 | | 0.8011 | | −11.9 | 14.4 | |
| 0.3080 | | −11.3 | 10.3 | | 0.8520 | | −9.8 | 12.4 | |
| 0.3485 | | −12.5 | 11.6 | | 0.9015 | | −7.3 | 9.6 | |
| 0.4006 | | −13.5 | 12.8 | | 0.9516 | | −3.9 | | |
| 0.5048 | | −14.8 | 14.9 | | 1.0000 | 0.98 | 0.0 | 0.0 | |

Uncertainties are: $u(x_1) = 0.0001$; $u(\rho) = 1\text{kPa}$; $u(T) = 0.01\text{K}$; and the combined expanded uncertainties (0.95 level of confidence) are: $U_{rc}(\kappa_S^{\text{E}}) = 0.04\,|\kappa_S^{\text{E}}|$, $U_{rc}(c^{\text{E}}) = 0.030\,|c^{\text{E}}|$ and $U_{rc}(\alpha_p^{\text{E}}) = 0.005 \cdot 10^{-3}\,\text{K}^{-1}$.

TABLE 5

Excess refractive indices, $n_D$, and the corresponding excess values, $n_D^E$ of *N,N*-dimethylacetamide (1) + 2-alkanone(2) mixtures at temperature *T* and 0.1 MPa.

| $x_1$ | $n_D$ | $n_D^E$ | $x_1$ | $n_D$ | $n_D^E$ |
|---|---|---|---|---|---|
| | | *N,N*-dimethylacetamide (1) + 2-propanone (2) *T*/K= 293.15 | | | |
| 0.0000 | 1.35854 | 0.00000 | 0.5524 | 1.40923 | 0.00188 |
| 0.0558 | 1.36466 | 0.00047 | 0.5974 | 1.41257 | 0.00184 |
| 0.1019 | 1.36946 | 0.00074 | 0.6502 | 1.41638 | 0.00179 |
| 0.1470 | 1.37405 | 0.00102 | 0.7022 | 1.41997 | 0.00167 |
| 0.2021 | 1.37947 | 0.00132 | 0.7502 | 1.42313 | 0.00148 |
| 0.2523 | 1.38417 | 0.00149 | 0.8004 | 1.42638 | 0.00131 |
| 0.2991 | 1.38840 | 0.00161 | 0.8538 | 1.42966 | 0.00105 |
| 0.3525 | 1.39306 | 0.00170 | 0.8980 | 1.43235 | 0.00087 |
| 0.3991 | 1.39704 | 0.00181 | 0.9468 | 1.43514 | 0.00057 |
| 0.4452 | 1.40083 | 0.00185 | 1.0000 | 1.43787 | 0.00000 |
| 0.4994 | 1.40517 | 0.00190 | | | |
| | | *N,N*-dimethylacetamide (1) + 2-propanone (2) *T*/K= 298.15 | | | |
| 0.0000 | 1.35386 | 0.00000 | 0.5547 | 1.40776 | 0.00254 |
| 0.0549 | 1.36261 | 0.00112 | 0.6113 | 1.41169 | 0.00220 |
| 0.1023 | 1.36813 | 0.00195 | 0.6531 | 1.41456 | 0.00199 |
| 0.1573 | 1.37400 | 0.00254 | 0.7026 | 1.41776 | 0.00163 |
| 0.2021 | 1.37850 | 0.00286 | 0.7528 | 1.42098 | 0.00133 |
| 0.2522 | 1.38334 | 0.00315 | 0.7985 | 1.42383 | 0.00104 |
| 0.2989 | 1.38759 | 0.00327 | 0.8470 | 1.42680 | 0.00076 |
| 0.3512 | 1.39211 | 0.00328 | 0.9009 | 1.43002 | 0.00045 |
| 0.3995 | 1.39607 | 0.00319 | 0.9441 | 1.43258 | 0.00024 |
| 0.4457 | 1.39965 | 0.00300 | 1.0000 | 1.43584 | 0.00000 |
| 0.4963 | 1.40353 | 0.00284 | | | |

TABLE 5 (continued)

*N,N*-dimethylacetamide (1) + 2-propanone (2) *T*/K= 303.15

| | | | | | |
|---|---|---|---|---|---|
| 0.0000 | 1.35323 | 0.00000 | 0.5524 | 1.40488 | 0.00222 |
| 0.0558 | 1.35961 | 0.00067 | 0.5974 | 1.40820 | 0.00211 |
| 0.1019 | 1.36462 | 0.00111 | 0.6502 | 1.41197 | 0.00195 |
| 0.1470 | 1.36938 | 0.00150 | 0.7022 | 1.41546 | 0.00167 |
| 0.2021 | 1.37479 | 0.00174 | 0.7502 | 1.41864 | 0.00145 |
| 0.2523 | 1.37966 | 0.00202 | 0.8004 | 1.42186 | 0.00119 |
| 0.2991 | 1.38411 | 0.00230 | 0.8538 | 1.42521 | 0.00094 |
| 0.3525 | 1.38875 | 0.00231 | 0.8980 | 1.42797 | 0.00078 |
| 0.3991 | 1.39281 | 0.00244 | 0.9468 | 1.43070 | 0.00036 |
| 0.4452 | 1.39655 | 0.00238 | 1.0000 | 1.43370 | 0.00000 |
| 0.4994 | 1.40081 | 0.00229 | | | |

*N,N*-dimethylacetamide (1) + 2-butanone (2) *T*/K= 293.15

| | | | | | |
|---|---|---|---|---|---|
| 0.0000 | 1.37839 | 0.00000 | 0.5510 | 1.41334 | 0.00123 |
| 0.0585 | 1.38245 | 0.00038 | 0.5993 | 1.41615 | 0.00117 |
| 0.1098 | 1.38591 | 0.00063 | 0.6556 | 1.41937 | 0.00108 |
| 0.1550 | 1.38890 | 0.00081 | 0.7049 | 1.42219 | 0.00101 |
| 0.2085 | 1.39238 | 0.00099 | 0.7570 | 1.42511 | 0.00090 |
| 0.2537 | 1.39528 | 0.00111 | 0.8008 | 1.42752 | 0.00077 |
| 0.3022 | 1.39832 | 0.00118 | 0.8488 | 1.43016 | 0.00065 |
| 0.3501 | 1.40127 | 0.00121 | 0.9052 | 1.43319 | 0.00044 |
| 0.3997 | 1.40432 | 0.00126 | 0.9469 | 1.43532 | 0.00020 |
| 0.4566 | 1.40775 | 0.00127 | 1.0000 | 1.43814 | 0.00000 |
| 0.5019 | 1.41045 | 0.00126 | | | |

*N,N*-dimethylacetamide (1) + 2-butanone (2) *T*/K= 298.15

| | | | | | |
|---|---|---|---|---|---|
| 0.0000 | 1.37612 | 0.00000 | 0.5412 | 1.41035 | 0.00099 |
| 0.0551 | 1.37995 | 0.00035 | 0.6016 | 1.41383 | 0.00088 |
| 0.1073 | 1.38349 | 0.00062 | 0.6560 | 1.41693 | 0.00076 |
| 0.1536 | 1.38656 | 0.00081 | 0.7067 | 1.41979 | 0.00064 |

TABLE 5 (continued)

| | | | | | |
|---|---|---|---|---|---|
| 0.2041 | 1.38982 | 0.00093 | 0.7524 | 1.42236 | 0.00054 |
| 0.2536 | 1.39299 | 0.00105 | 0.8052 | 1.42531 | 0.00041 |
| 0.2936 | 1.39551 | 0.00112 | 0.8521 | 1.42792 | 0.00031 |
| 0.3473 | 1.39882 | 0.00115 | 0.9056 | 1.43088 | 0.00018 |
| 0.4052 | 1.40231 | 0.00112 | 0.9459 | 1.43311 | 0.00011 |
| 0.4522 | 1.40512 | 0.00110 | 1.0000 | 1.43609 | 0.00000 |
| 0.4958 | 1.40768 | 0.00103 | | | |
| *N,N*-dimethylacetamide (1) + 2-butanone (2) $T$/K= 303.15 | | | | | |
| 0.0000 | 1.37351 | 0.00000 | 0.5948 | 1.41061 | 0.00057 |
| 0.0672 | 1.37816 | 0.00041 | 0.6587 | 1.41429 | 0.00046 |
| 0.1269 | 1.38214 | 0.00064 | 0.6964 | 1.41644 | 0.00038 |
| 0.1810 | 1.38566 | 0.00079 | 0.7447 | 1.41921 | 0.00031 |
| 0.2406 | 1.38944 | 0.00088 | 0.7940 | 1.42201 | 0.00023 |
| 0.2933 | 1.39272 | 0.00092 | 0.8307 | 1.42411 | 0.00019 |
| 0.3526 | 1.39633 | 0.00089 | 0.8727 | 1.42649 | 0.00013 |
| 0.3955 | 1.39891 | 0.00086 | 0.9140 | 1.42882 | 0.00008 |
| 0.4485 | 1.40206 | 0.00080 | 0.9545 | 1.43111 | 0.00004 |
| 0.5031 | 1.40527 | 0.00072 | 1.0000 | 1.43368 | 0.00000 |
| 0.5410 | 1.40749 | 0.00066 | | | |
| *N,N*-dimethylacetamide (1) + 2-pentanone (2) $T$/K= 293.15 | | | | | |
| 0.0000 | 1.39016 | 0.00000 | 0.5550 | 1.41614 | 0.00069 |
| 0.0530 | 1.39259 | 0.00016 | 0.6075 | 1.41871 | 0.00068 |
| 0.1061 | 1.39501 | 0.00027 | 0.6513 | 1.42087 | 0.00066 |
| 0.1551 | 1.39726 | 0.00037 | 0.6982 | 1.42319 | 0.00062 |
| 0.2004 | 1.39936 | 0.00046 | 0.7407 | 1.42532 | 0.00060 |
| 0.2552 | 1.40190 | 0.00053 | 0.7992 | 1.42826 | 0.00052 |
| 0.3070 | 1.40431 | 0.00058 | 0.8532 | 1.43100 | 0.00043 |
| 0.3637 | 1.40698 | 0.00064 | 0.9004 | 1.43339 | 0.00032 |
| 0.4083 | 1.40909 | 0.00066 | 0.9439 | 1.43560 | 0.00020 |

TABLE 5 (continued)

| | | | | | |
|---|---|---|---|---|---|
| 0.4566 | 1.41140 | 0.00069 | 1.0000 | 1.43845 | 0.00000 |
| 0.5013 | 1.41354 | 0.00069 | | | |
| *N,N*-dimethylacetamide (1) + 2-pentanone (2) *T*/K= 298.15 | | | | | |
| 0.0000 | 1.38773 | 0.00000 | 0.5497 | 1.41368 | 0.00083 |
| 0.0540 | 1.39022 | 0.00017 | 0.6041 | 1.41634 | 0.00081 |
| 0.1037 | 1.39252 | 0.00030 | 0.6373 | 1.41799 | 0.00081 |
| 0.1497 | 1.39466 | 0.00042 | 0.6983 | 1.42102 | 0.00077 |
| 0.2016 | 1.39707 | 0.00052 | 0.7431 | 1.42326 | 0.00071 |
| 0.2553 | 1.39958 | 0.00060 | 0.7979 | 1.42601 | 0.00063 |
| 0.2990 | 1.40163 | 0.00066 | 0.8446 | 1.42837 | 0.00053 |
| 0.3427 | 1.40371 | 0.00072 | 0.8957 | 1.43095 | 0.00040 |
| 0.3951 | 1.40621 | 0.00077 | 0.9392 | 1.43314 | 0.00025 |
| 0.4452 | 1.40862 | 0.00081 | 1.0000 | 1.43621 | 0.00000 |
| 0.4986 | 1.41120 | 0.00083 | | | |
| *N,N*-dimethylacetamide (1) + 2-pentanone (2) *T*/K= 303.15 | | | | | |
| 0.0000 | 1.38530 | 0.00000 | 0.5495 | 1.41121 | 0.00087 |
| 0.0535 | 1.38778 | 0.00019 | 0.5945 | 1.41343 | 0.00088 |
| 0.0989 | 1.38990 | 0.00034 | 0.6408 | 1.41571 | 0.00086 |
| 0.1528 | 1.39242 | 0.00049 | 0.6965 | 1.41847 | 0.00082 |
| 0.2015 | 1.39467 | 0.00058 | 0.7468 | 1.42099 | 0.00077 |
| 0.2532 | 1.39708 | 0.00066 | 0.7993 | 1.42364 | 0.00071 |
| 0.3051 | 1.39952 | 0.00074 | 0.8450 | 1.42596 | 0.00063 |
| 0.3498 | 1.40164 | 0.00080 | 0.9015 | 1.42877 | 0.00044 |
| 0.3969 | 1.40386 | 0.00082 | 0.9476 | 1.43108 | 0.00027 |
| 0.4491 | 1.40636 | 0.00085 | 1.0000 | 1.43367 | 0.00000 |
| 0.4998 | 1.40880 | 0.00087 | | | |

TABLE 5 (continued)

| | | | | | |
|---|---|---|---|---|---|
| | *N,N*-dimethylacetamide (1) + 2-heptanone (2) $T$/K= 298.15 | | | | |
| 0.0000 | 1.40691 | 0.00000 | 0.5571 | 1.42057 | 0.00031 |
| 0.0578 | 1.40811 | 0.00005 | 0.6080 | 1.42209 | 0.00030 |
| 0.1069 | 1.40917 | 0.00010 | 0.6492 | 1.42337 | 0.00030 |
| 0.1570 | 1.41027 | 0.00013 | 0.6981 | 1.42495 | 0.00029 |
| 0.2088 | 1.41146 | 0.00017 | 0.7499 | 1.42668 | 0.00027 |
| 0.2529 | 1.41250 | 0.00020 | 0.8011 | 1.42846 | 0.00024 |
| 0.3080 | 1.41384 | 0.00023 | 0.8520 | 1.43031 | 0.00020 |
| 0.3485 | 1.41485 | 0.00024 | 0.9015 | 1.43218 | 0.00014 |
| 0.4006 | 1.41621 | 0.00028 | 0.9516 | 1.43416 | 0.00008 |
| 0.4484 | 1.41749 | 0.00029 | 1.0000 | 1.43615 | 0.00000 |
| 0.5048 | 1.41905 | 0.00030 | | | |

Uncertainties are: $u(T) = 0.02\text{K}$ ; $u(p) = 1\text{kPa}$ ; $u(x_1) = 0.0001$; $u(n_\text{D}) = 0.00002$ ; the combined expanded uncertainty (0.95 level of confidence) is $U_{rc}(n_\text{D}{}^\text{E}) = 0.02 n_\text{D}{}^\text{E}$ .

TABLE 6

Coefficients $A_i$ and standard deviations, $\sigma(F^E)$ Eq. (9) for representation of the $F^{E,a}$ property at 298.15 K and 0.1 MPa for *N,N*-dimethylacetamide (1) + 2-alkanone (2) systems by Eq. (7).

| System | $T$ / K | Property $F^E$ | $A_0$ | $A_1$ | $A_2$ | $A_3$ | $\sigma(F^E)$ |
|---|---|---|---|---|---|---|---|
| DMA + 2-propanone | 293.15 | $V_m^E$ | −1.484 | 0.345 | −0.137 | | 0.003 |
| | | $n_D^E$ | 0.00751 | 0.00033 | 0.00193 | | 0.00004 |
| | 298.15 | $V_m^E$ | −1.561 | 0.391 | −0.15 | | 0.004 |
| | | $c^E$ | 225.3 | 15.4 | 7.3 | | 0.3 |
| | | $\kappa_S^E$ | −248.0 | 94.2 | −30.4 | | 0.2 |
| | | $\alpha_p^E$ | −0.2047 | 0.0467 | −0.0786 | 0.0387 | 0.0007 |
| | | $n_D^E$ | 0.01127 | −0.00960 | 0.00269 | | 0.00003 |
| | 303.15 | $V_m^E$ | −1.668 | 0.369 | −0.185 | | 0.004 |
| | | $n_D^E$ | 0.00938 | −0.00300 | | | 0.00005 |
| DMA + 2-butanone | 293.15 | $V_m^E$ | −0.928 | 0.108 | −0.101 | | 0.003 |
| | | $n_D^E$ | 0.005025 | −0.000981 | 0.001119 | | 0.000017 |
| | 298.15 | $V_m^E$ | −0.9996 | 0.1589 | −0.1164 | | 0.0018 |
| | | $c^E$ | 158.34 | 34.09 | 11.00 | | 0.08 |
| | | $\kappa_S^E$ | −171.37 | 29.17 | −8.01 | 4.27 | 0.06 |
| | | $\alpha_p^E$ | −0.1117 | 0.0302 | 0.0988 | | 0.0004 |
| | | $n_D^E$ | 0.004172 | −0.00268 | 0.00028 | | 0.00001 |
| | 303.15 | $V_m^E$ | −1.037 | 0.140 | | | 0.003 |
| | | $n_D^E$ | 0.002905 | −0.003135 | 0.001074 | | 0.000005 |
| DMA + 2-pentanone | 293.15 | $V_m^E$ | −0.7277 | 0.0294 | | | 0.0019 |
| | | $n_D^E$ | 0.002760 | 0.000336 | 0.000766 | | 0.000008 |
| | 298.15 | $V_m^E$ | −0.8137 | 0.0781 | −0.0833 | −0.1273 | 0.0017 |
| | | $c^E$ | 126.3 | 48.5 | 20.3 | | 0.3 |
| | | $\kappa_S^E$ | −136.53 | −1.91 | −3.18 | | 0.14 |
| | | $\alpha_p^E$ | −0.1218 | 0.0104 | −0.1050 | −0.0352 | 0.0007 |
| | | $n_D^E$ | 0.003295 | 0.000575 | 0.000730 | | 0.000006 |

TABLE 6 (continued)

| | | | | | | | |
|---|---|---|---|---|---|---|---|
| | 303.15 | $V_m^E$ | −0.8703 | 0.0553 | −0.1347 | | 0.0016 |
| | | $n_D^E$ | 0.003474 | 0.000629 | 0.001500 | | 0.000012 |
| DMA + 2-heptanone | 298.15 | $V_m^E$ | −0.046 | 0.007 | | | 0.005 |
| | | $c^E$ | 59.36 | 35.51 | 23.12 | 7.62 | 0.08 |
| | | $\kappa_S^E$ | −59.44 | −19.44 | −10.04 | | 0.08 |
| | | $n_D^E$ | 0.001198 | 0.000388 | 0.000172 | | 0.000005 |

$F^E = V_m^E$, units: $cm^3 mol^{-1}$; $F^E = c^E$, units: $ms^{-1}$; $F^E = \kappa_S^E$ units: $TPa^{-1}$; $F^E = \alpha_p^E$, units: $10^{-3} K^{-1}$.

TABLE 7

Molar excess enthalpies, $H_m^{E}$, at equimolar composition and 298.15 K for tertiary amide(1) + *n*-alkanone(2) mixtures. The interaction parameters, $X_{12}$, calculated from $H_m^{E}$ values at equimolar composition and the interactional contribution, $H_{m,int}^{E}$, to $H_m^{E}$ are also included.

| System | $H_m^{E}$ /J·mol$^{-1}$ | $X_{12}$ /J·cm$^{-3}$ | $H_{m,int}^{E}$ /J·mol$^{-1}$ | $\sigma_r(H_m^{E})$ [a] |
|---|---|---|---|---|
| DMF + 2-propanone | 34[16] | 3.97 | 46 | 0.050 |
| DMF + 2-butanone | 143[17] | 9.35 | 117 | 0.071 |
| DMF + 2-pentanone | 206[17] | 11.93 | 160 | 0.082 |
| DMF + 3-pentanone | 257[17] | 14.84 | 199 | 0.006 |
| NMP + 2-butanone | – 165[14] | – 6.14 | – 93 | 0.487 |
| NMP + 4-heptanone | 226[15] | 10.42 | 186 | 0.266 |
| NMP + 5-nonanone | 362[15] | 15.00 | 288 | 0.352 |

[a]relative standard deviation (eq. 15)

TABLE 8

Excess molar volumes, $V_{\mathrm{m}}^{\mathrm{E}}$, and isochoric excess molar internal energies, $U_{V,\mathrm{m}}^{\mathrm{E}}$, for tertiary amide(1) + *n*-alkanone(2) mixtures at 298.15 K and equimolar composition. The contributions to $V_{\mathrm{m}}^{\mathrm{E}}$ from the *P** and curvature terms, according to the Prigogine-Flory-Patterson model, and the equation of state contribution to $H_{\mathrm{m}}^{\mathrm{E}}$ (Eos, see equation 19) are also included.

| *n*-Alkanone | $V_{\mathrm{m}}^{\mathrm{E}}$ /cm$^3$·mol$^{-1}$ | | | | $\left\|(V_{\mathrm{m,cur}}^{\mathrm{E}}+V_{\mathrm{m,P^*}}^{\mathrm{E}})/V_{\mathrm{m,exp}}^{\mathrm{E}}\right\|$ | Eos/ J·mol$^{-1}$ | $U_{\mathrm{V,m}}^{\mathrm{E}}$ /J·mol$^{-1}$[a] |
|---|---|---|---|---|---|---|---|
| | Exp. | Flory | Curv. term | *P** term | | | |
| | | | | DMF + *n*-alkanone | | | |
| 2-propanone | − 0.4313[11] | − 0.345 | − 0.154 | − 0.235 | 0.90 | − 167 | 204 |
| 2-butanone | − 0.3115[11] | − 0.192 | − 0.082 | − 0.212 | 0.94 | − 119 | 262 |
| 2-pentanone | − 0.2510[11] | − 0.088 | − 0.038 | − 0.188 | 0.90 | − 97 | 303 |
| 3-pentanone | | | − 0.038 | − 0.171 | | | |
| 2-heptanone | − 0.0503[11] | | − 0.003 | − 0.066 | 1.37 | − 19 | |
| | | | | DMA + *n*-alkanone | | | |
| 2-propanone | − 0.3903[b] | | − 0.192 | − 0.222 | 1.06 | − 150 | |
| 2-butanone | − 0.2499[b] | | − 0.109 | − 0.215 | 1.30 | − 95 | |
| 2-pentanone | − 0.2034[b] | | − 0.056 | − 0.210 | 1.31 | − 75 | |
| 2-heptanone | − 0.0115[b] | | − 0.009 | − 0.105 | 9.91 | − 4 | |
| | | | | NMP + *n*-alkanone | | | |
| 2-propanone | − 0.7190[13] | | − 0.349 | − 0.310 | 0.92 | − 273 | |
| 2-butanone | − 0.6010[13] | − 0.638 | − 0.242 | − 0.332 | 0.96 | − 224 | 59 |
| 2-pentanone | − 0.5728[13] | | − 0.166 | − 0.370 | 0.94 | − 212 | |
| 3-pentanone | − 0.4975[13] | | − 0.165 | − 0.334 | 1.00 | − 187 | |
| 2-hexanone | − 0.4678[13] | | − 0.106 | − 0.319 | 0.90 | − 174 | |
| 4-heptanone | − 0.3910[13] | − 0.233 | − 0.072 | − 0.276 | 0.89 | − 146 | 372 |
| 5-nonanone | − 0.2143[13] | − 0.004 | − 0.024 | − 0.183 | 0.97 | − 80 | 442 |

[a]eq. (19); [b]this work;

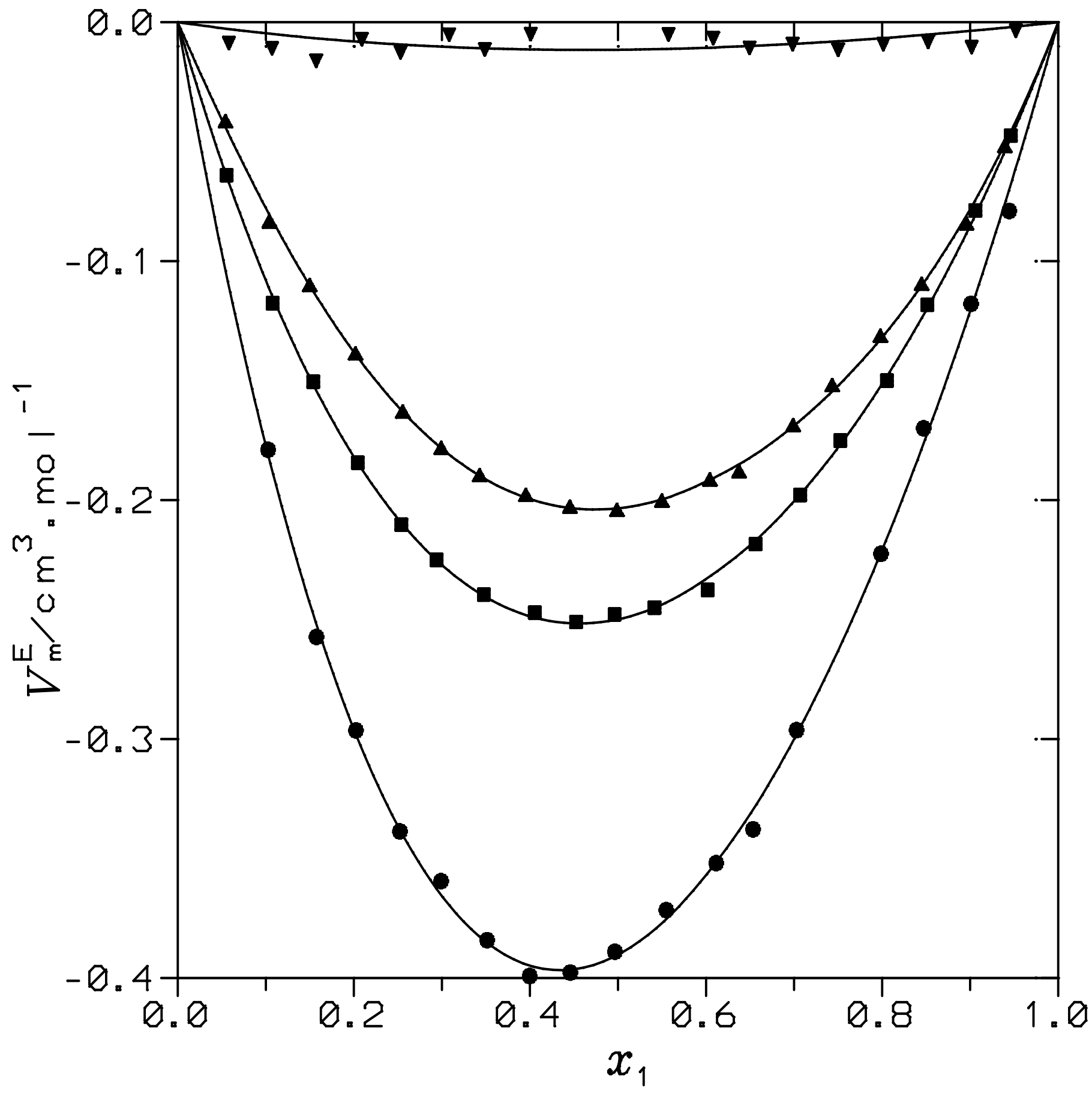

Figure 1 Excess molar volumes, $V_m^E$, for DMA (1) + 2-alkanone (2) systems at atmospheric pressure and 298.15 K. Full symbols, experimental values (this work): (●), 2-propanone; (■), 2-butanone; (▲), 2-pentanone, (♦), 2-heptanone. Solid lines, calculations with equation (8) using coefficients from Table 6.

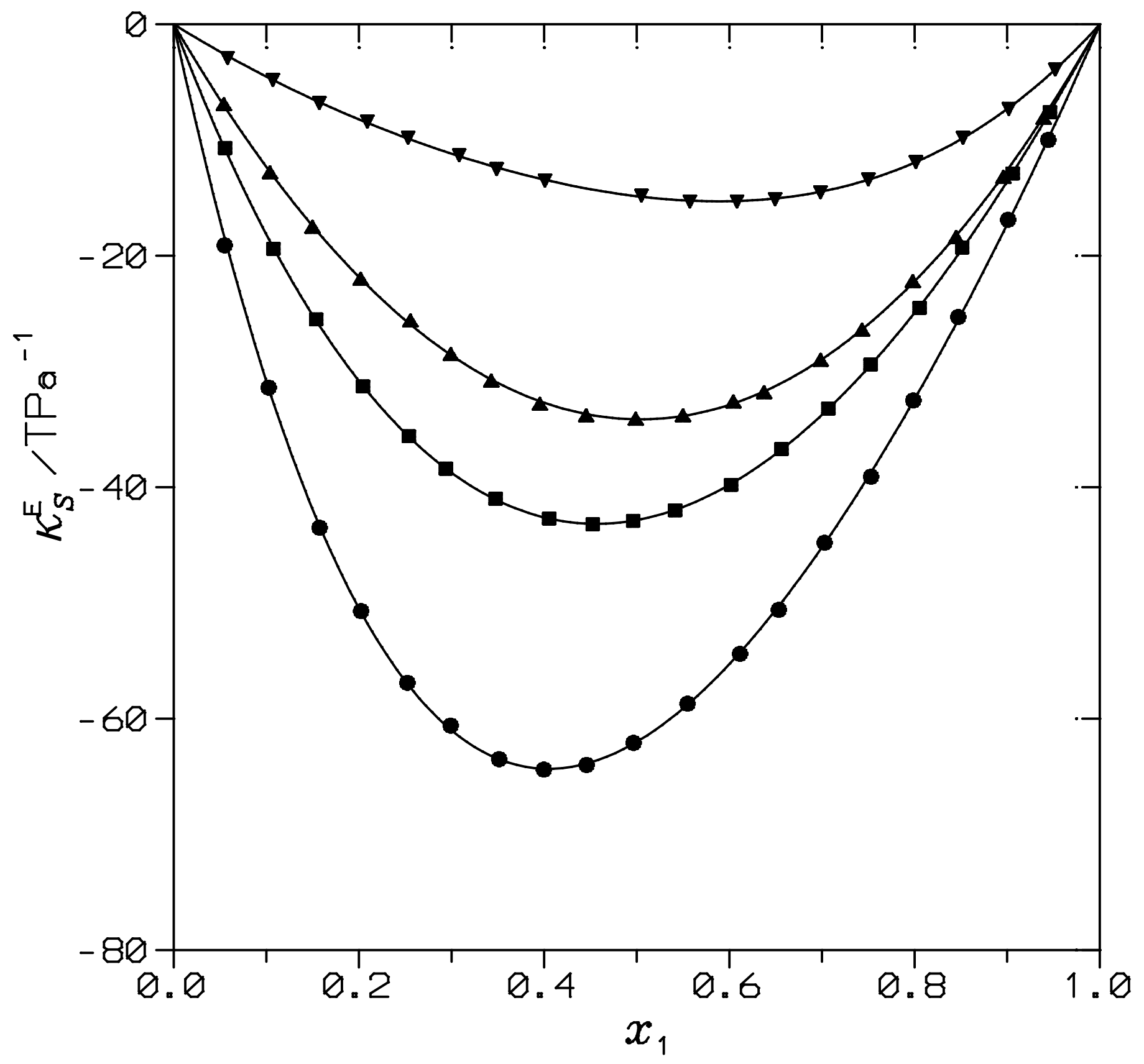

Figure 2 Excess isentropic compressibilities, $\kappa_S^{\mathrm{E}}$, for DMA (1) + 2-alkanone (2) systems at atmospheric pressure and 298.15 K. Full symbols, experimental values (this work): (●), 2-propanone; (■), 2-butanone; (▲), 2-pentanone, (♦), 2-heptanone. Solid lines, calculations with equation (8) using coefficients from Table 6.

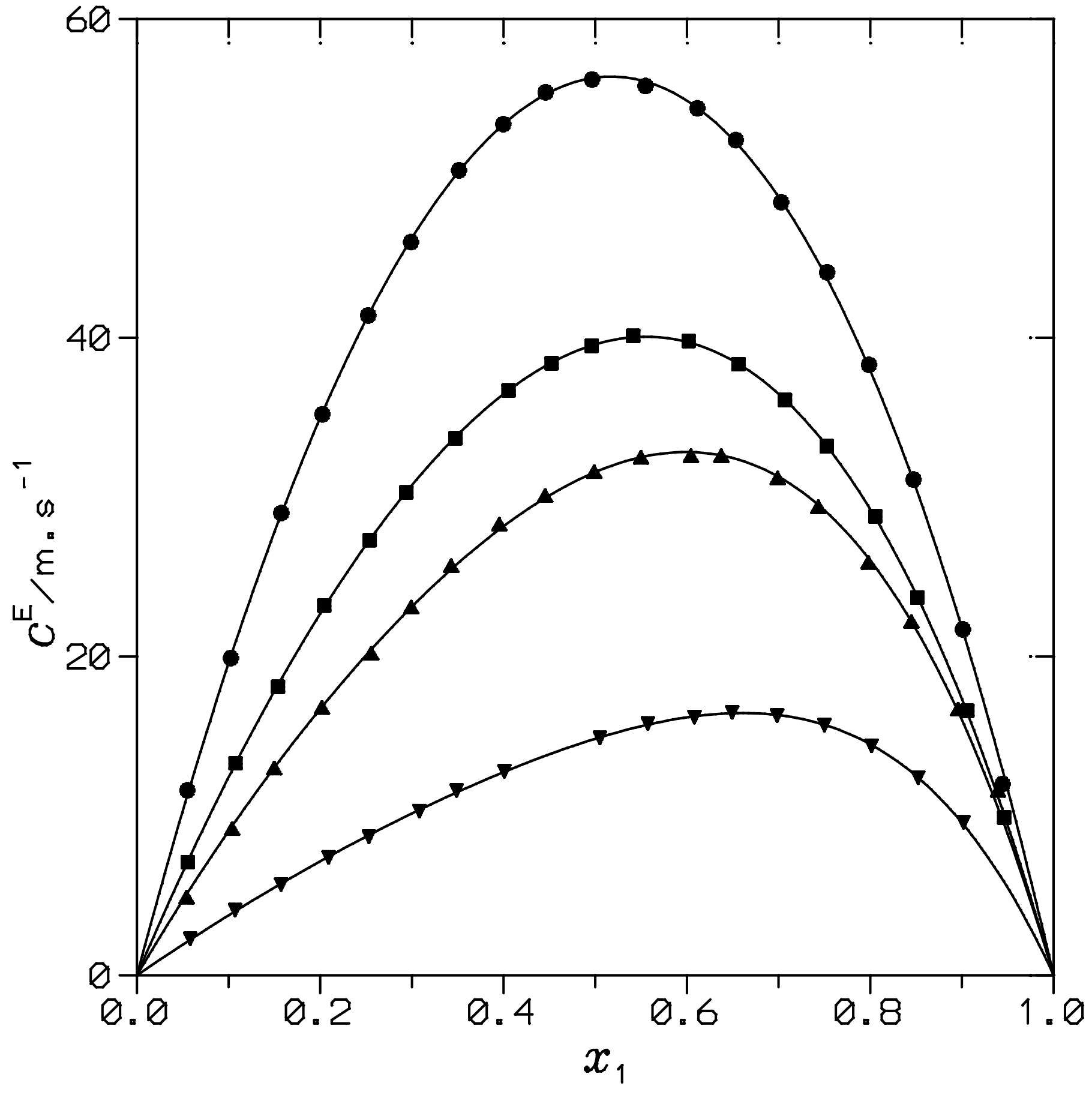

Figure 3 Excess speeds of sound, $c^{E}$, for DMA (1) + 2-alkanone (2) systems at atmospheric pressure and 298.15 K. Full symbols, experimental values (this work): (●), 2-propanone; (■), 2-butanone; (▲), 2-pentanone, (♦), 2-heptanone. Solid lines, calculations with equation (8) using coefficients from Table 6.

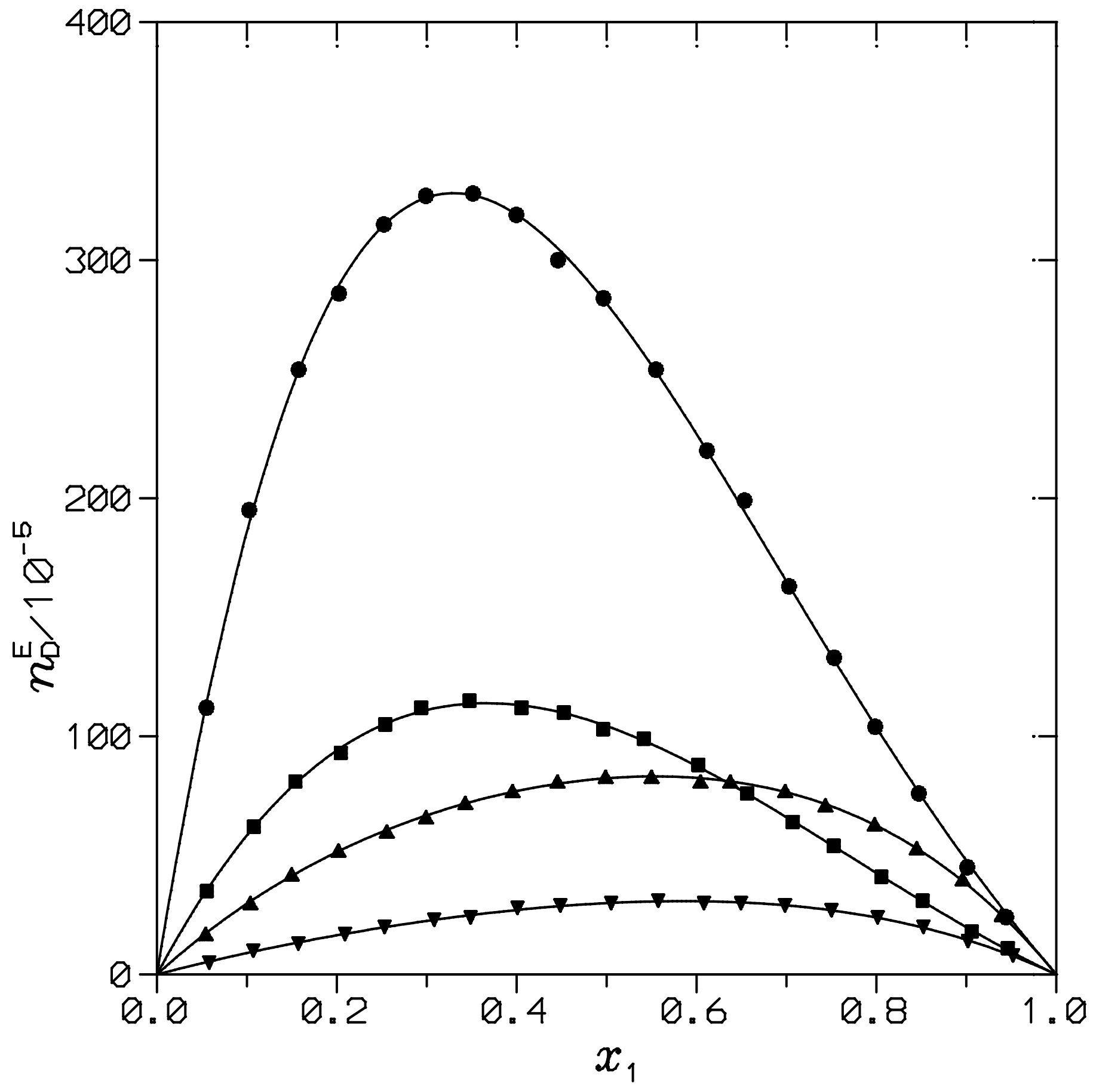

Figure 4 Excess refractive indices, $n_D^E$, for DMA (1) + 2-alkanone (2) systems at atmospheric pressure and 298.15 K. Full symbols, experimental values (this work): (●), 2-propanone; (■), 2-butanone; (▲), 2-pentanone, (♦), 2-heptanone. Solid lines, calculations with equation (8) using coefficients from Table 6.

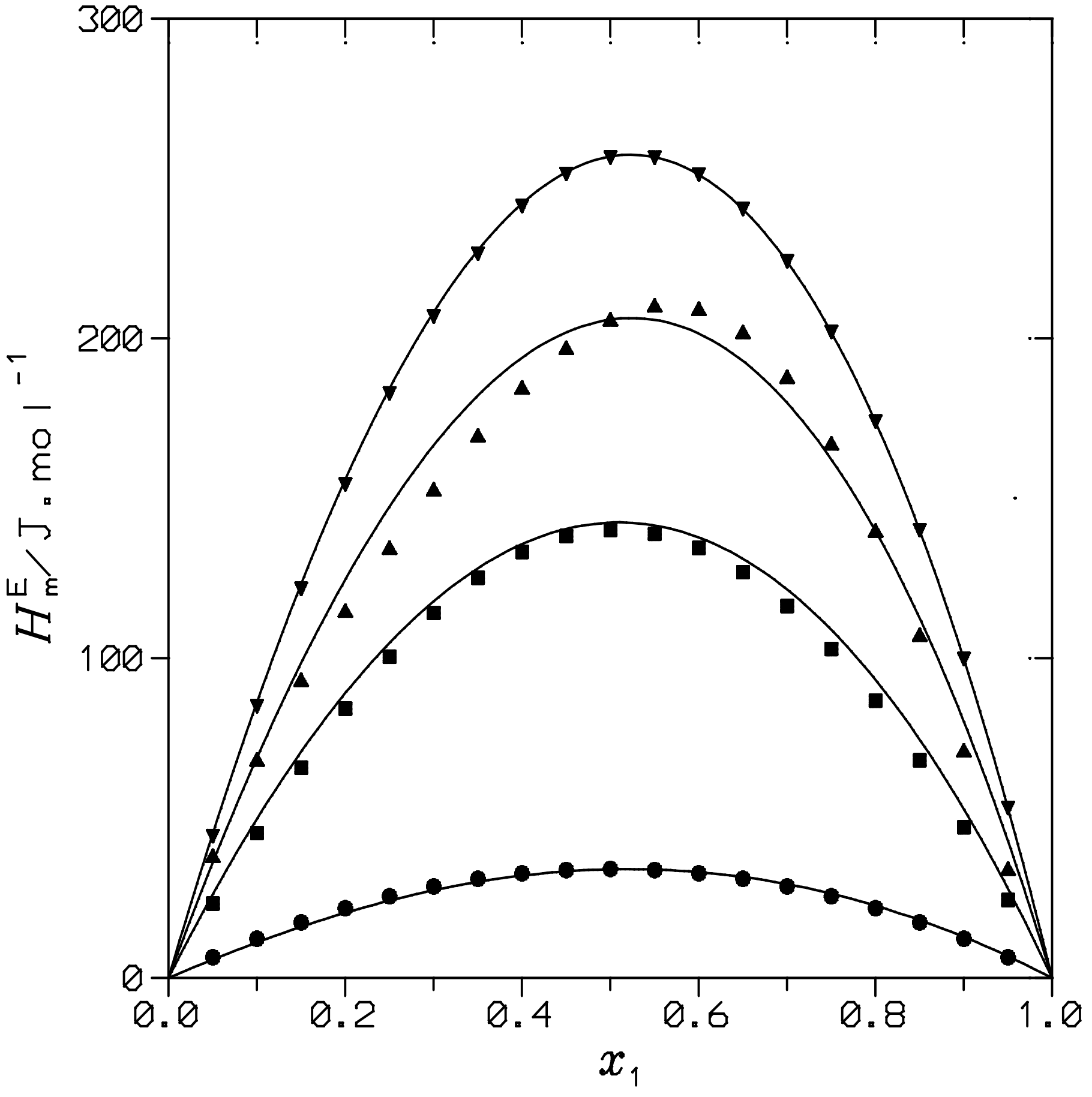

Figure 5 Excess molar enthalpies, $H^{\mathrm{E}}_{\mathrm{m}}$, for DMF (1) + 2-alkanone (2) systems at atmospheric pressure and 298.15 K. Full symbols, experimental results: (●), 2-propanone [16]; (■), 2-butanone [17]; (▲), 2-pentanone [17], (♦), 3-pentanone [17]. Solid lines, results from the Flory model using interaction parameters listed in Table 7.

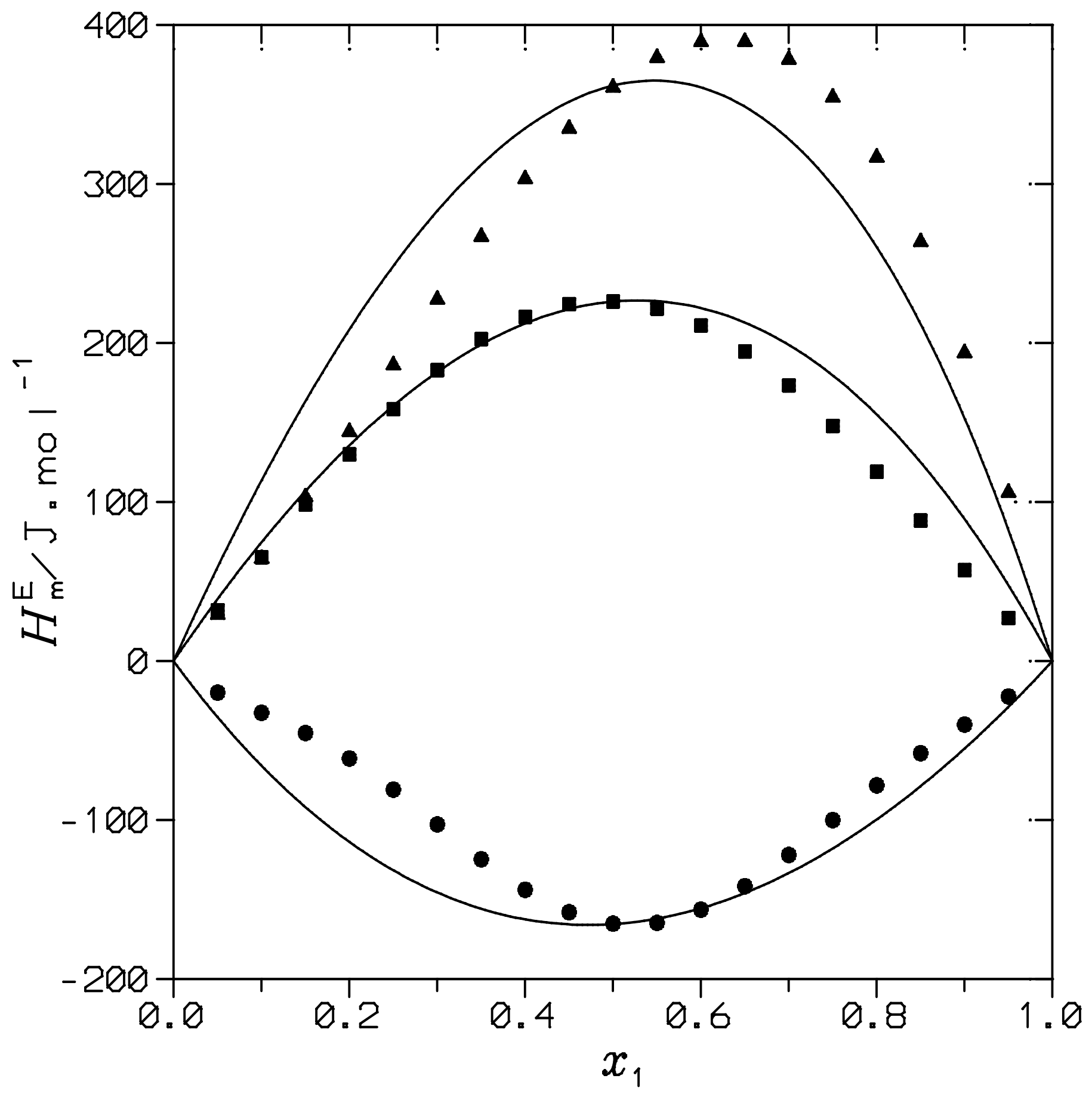

Figure 6 Excess molar enthalpies, $H_m^E$, for NMP (1) + 2-alkanone (2) systems at atmospheric pressure and 298.15 K. Full symbols, experimental results: (●), 2-butanone [14]; (■), 4-heptanone [15]; (▲), 5-nonanone [15]. Solid lines, results from the Flory model using interaction parameters listed in Table 7.

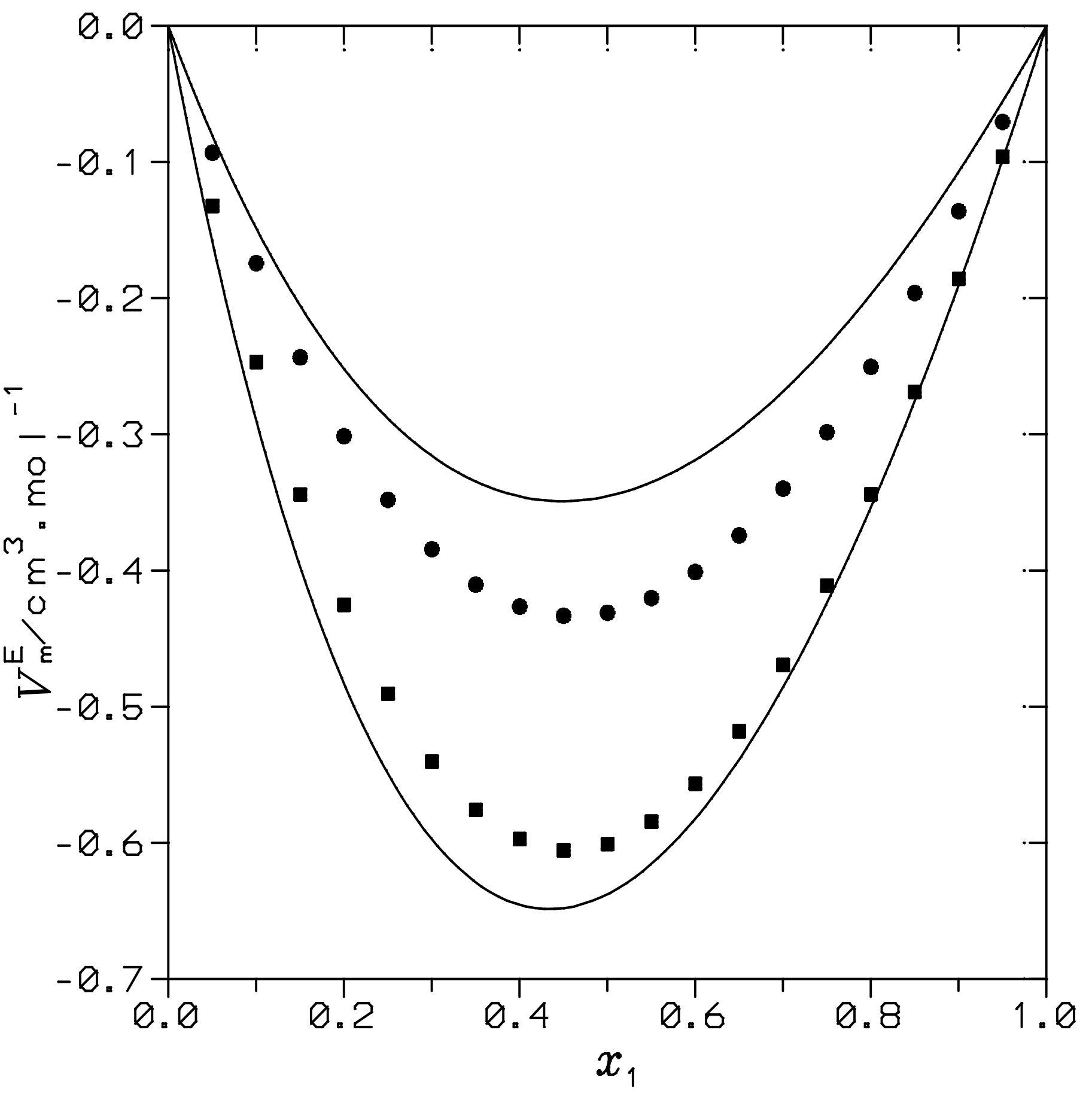

Figure 7 Excess molar volumes, $V_m^E$, for DMF (1) + 2-propanone (2) and NMP (1) + 2-butanone systems at atmospheric pressure and 298.15 K. Full symbols, experimental values: (●), DMF + 2-propanone [13]; (■), NMP + 2-butanone [15]. Solid lines, results from the Flory model using interaction parameters listed in Table 7.

# SUPPLEMENTARY MATERIAL

# Thermodynamics of amide + ketone mixtures. 2. Volumetric, speed of sound and refractive index data for *N,N*-dimethylacetamide + 2-alkanone systems at several temperatures. Application of Flory's model to tertiary amide + *n*-alkanone systems

ANA COBOS[(1)], JUAN ANTONIO GONZÁLEZ*[(1)], FERNANDO HEVIA,[(1)] ISAÍAS GARCÍA DE LA FUENTE[(1)] AND CRISTINA ALONSO TRISTÁN[(2)]

[(1)] Dpto. Ingeniería Electromecánica. Escuela Politécnica Superior. Avda. Cantabria s/n. 09006 Burgos, (Spain)

[(2)] G.E.T.E.F., Departamento de Física Aplicada, Facultad de Ciencias, Universidad de Valladolid, Paseo de Belén, 7, 47011 Valladolid, Spain,

*e-mail: jagl@termo.uva.es; Fax: +34-983-423136; Tel: +34-983-423757

TABLE S1

Molar volumes, $V_{m,i}$, thermal expansion coefficients, $\alpha_{p,i}$, isothermal compressibilities, $\kappa_{T,i}$ and Flory reduction parameters for volume, $V_{mi}^{*}$, and pressure, $p_i^{*}$, at 298.15 K and atmospheric pressure for pure compounds considered in this work.

| Compound | $V_{m,i}$ / cm$^3$mol$^{-1}$ | $\alpha_{p,i}$ / $10^{-3}$ K$^{-1}$ | $\kappa_{T,i}$ / TPa$^{-1}$ | $V_{mi}^{*}$ /cm$^3$·mol$^{-1}$ | $p_i^{*}$ / MPa |
|---|---|---|---|---|---|
| DMF | 77.43 | 1.01[13] | 660[13] | 61.95 | 712.6 |
| DMA | 93.04 | 0.98[a] | 657[a] | 74.81 | 687.7 |
| NMP | 96.43 | 0.84[15] | 532[93] | 79.48 | 690.3 |
| Acetone | 74.04 | 1.46[a] | 1316[a] | 55.46 | 589.3 |
| Butanone | 90.15 | 1.31[a] | 1172[a] | 68.92 | 570.0 |
| 2-Pentanone | 107.44 | 1.20[a] | 1098[a] | 83.46 | 539.8 |
| 3-Pentanone | 106.34 | 1.20[15] | 1073[15] | 82.60 | 552.5 |
| 2-Hexanone | 124.12 | 1.11[15] | 997[15] | 97.74 | 535.1 |
| 2-Heptanone | 140.76 | 1.06[a] | 967[a] | 111.72 | 518.7 |
| 4-Heptanone | 140.79 | 1.05[15] | 932[15] | 111.92 | 531.4 |
| 5-Nonanone | 174.02 | 0.95[15] | 838[15] | 140.63 | 517.4 |

[a]this work

TABLE S2

Partial excess molar enthalpies, $H_{m1}^{E,\infty}$, at 298.15 K for solute (1) + organic compound (2) mixtures, and enthalpy of the (amide + ketone) interaction, $\Delta H_{NCO\text{-}CO}$, Eq. (18), at 298.15 K for amide (1) + alkanone (2) systems.

| System | $H_{m1}^{E,\infty}$ / kJ mol$^{-1}$ | $\Delta H_{NCO\text{-}CO}$ / kJ mol$^{-1}$ |
|---|---|---|
| 2-propanone+heptane | 9.09[53] | |
| 2-butanone+heptane | 7.47[50] | |
| 2-pentanone+heptane | 6.35[50] | |
| 3-pentanone+heptane | 5.91[50] | |
| 2-heptanone+heptane | 5.57[51] | |
| 4-heptanone+heptane | 4.76[51] | |
| DMF+heptane | 17.1[47] | |
| NMP+heptane | 12.0[48] | |
| DMF+propanone | 0.15[16] | – 26.04 |
| DMF+2-butanone | 0.49[17] | – 24.08 |
| DMF+2-pentanone | 0.61[17] | – 22.84 |
| DMF+3-pentanone | 0.90[17] | – 22.11 |
| NMP+2-butanone | 0.54[14] | – 20.01 |
| NMP+4-heptanone | 0.61[15] | – 16.15 |
| NMP+5-nonanone | 0.55[15] | |

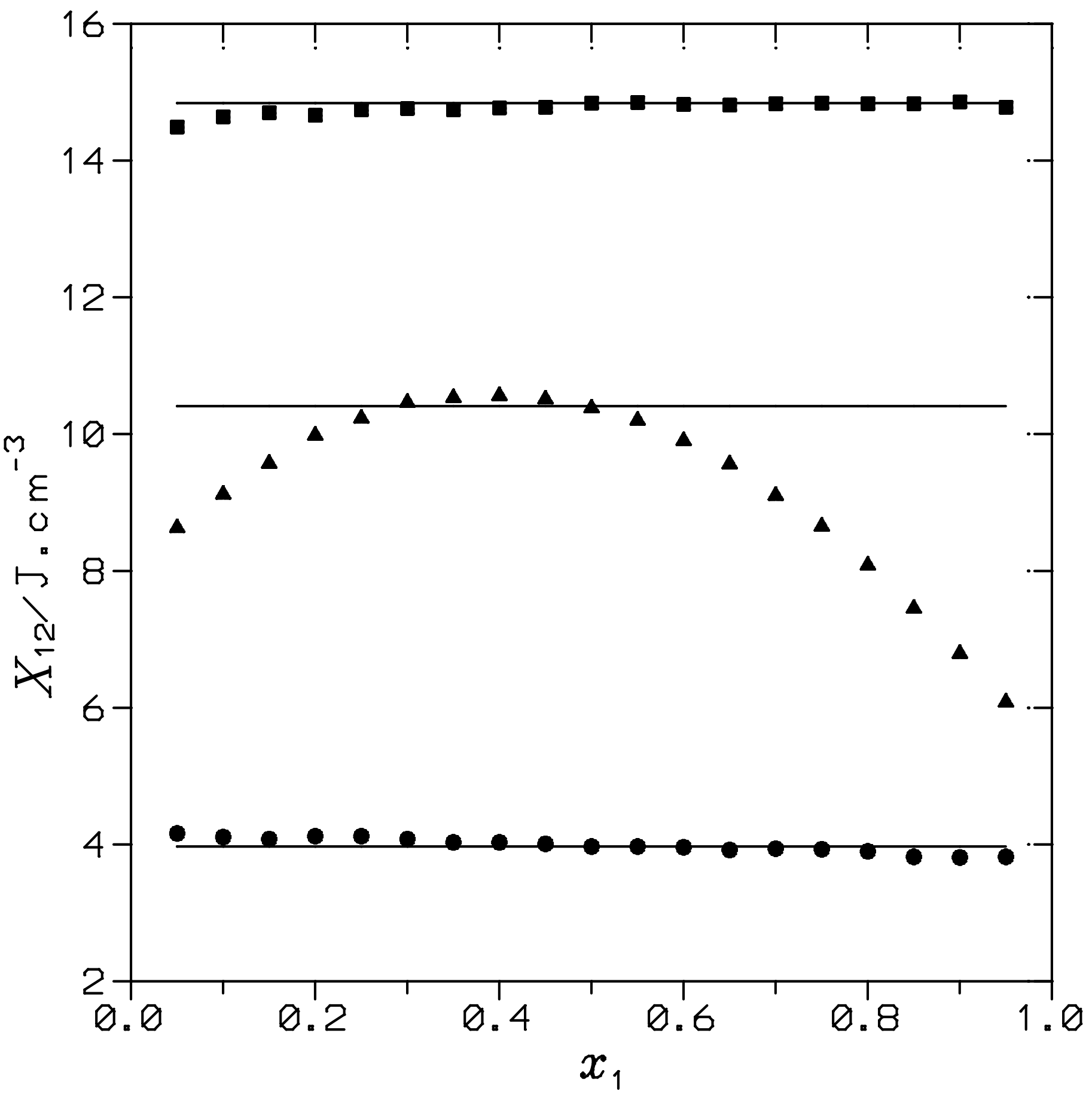

Figure S1 Flory interaction parameter, $X_{12}$ for amide(1) + *n*-alkanone(2) systems at atmospheric pressure and 298.15 K (this work): (●), DMF(1) + 2-propanone(2); (■), DMF(1) + 3-pentanone(2); (▲), NMP(1) + 4-heptanone(2).

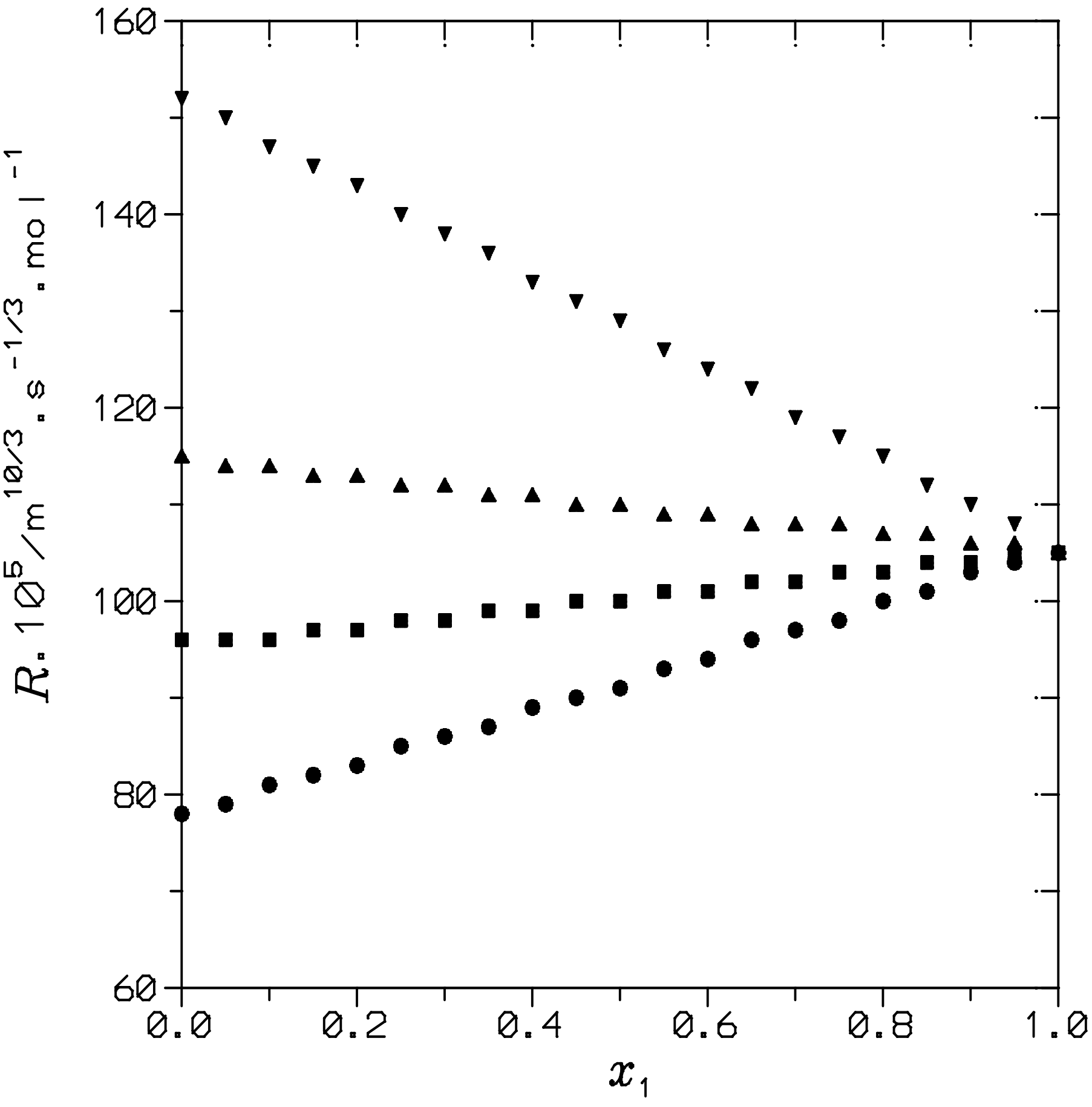

Figure S2  Rao's constant for DMA (1) + 2-alkanone (2) systems at atmospheric pressure and 298.15 K (this work): (●), 2-propanone; (■), 2-butanone; (▲), 2-pentanone, (♦), 2-heptanone.

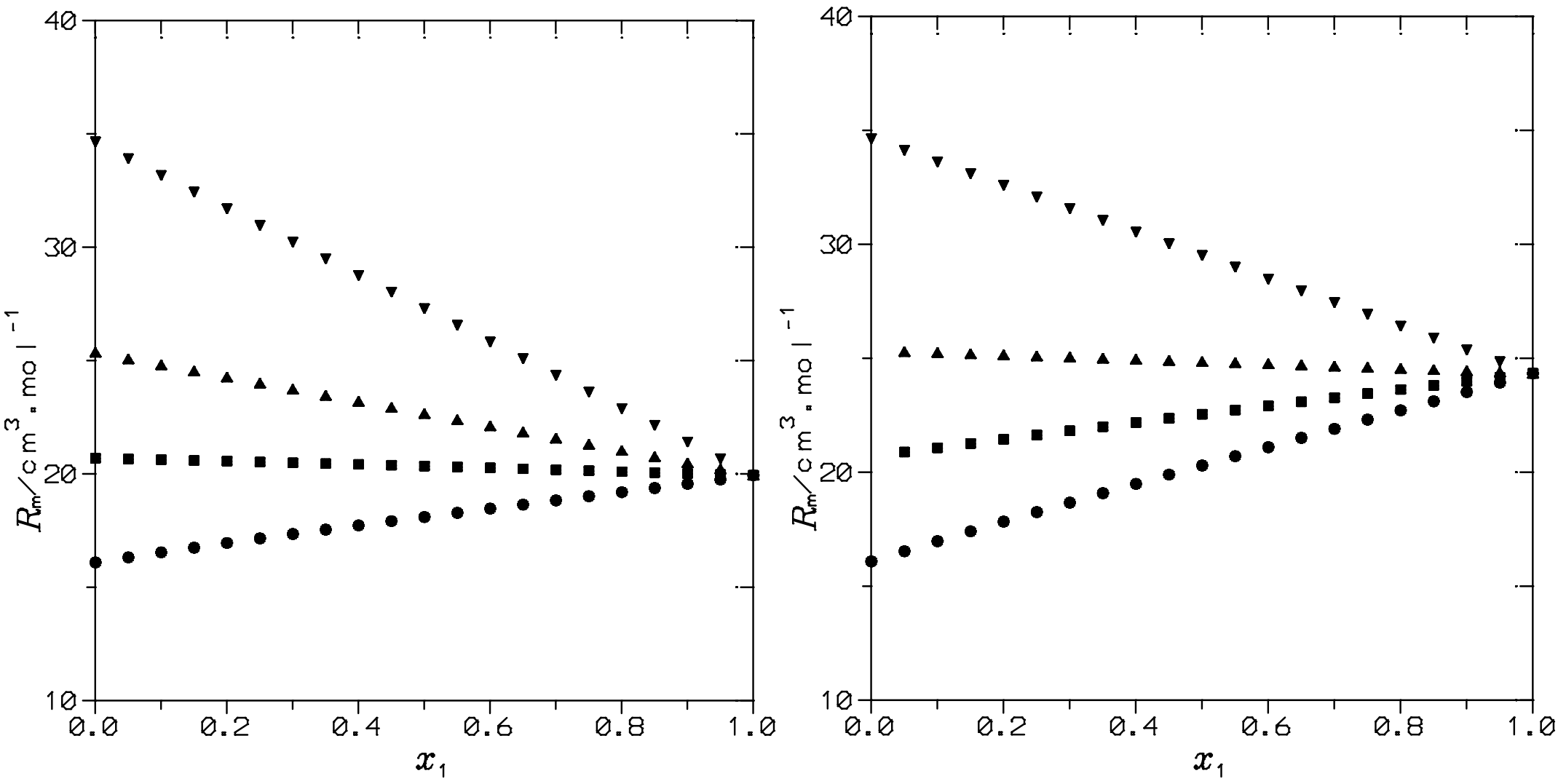

Figure S3 Molar refraction, $R_m$, for DMF (1) + 2-alkanone (2) (left figure) [13] and DMA (1) + 2-alkanone (2) (right figure) (this work) systems at atmospheric pressure and 298.15 K: (●), 2-propanone; (■), 2-butanone; (▲), 2-pentanone, (♦), 2-heptanone.